\documentclass[twocolumn,showpacs,superscriptaddress,email,floatfix,longbibliography,prl]{revtex4-2}
\usepackage{braket}
\usepackage{natbib}
\usepackage{xcolor}
\usepackage{amsmath}
\usepackage{array}
\usepackage{tabularx}
\usepackage{mathtools}
\usepackage{commath}
\usepackage{bbold}
\usepackage{graphicx}
\graphicspath{ {Images/} }
\usepackage{hyperref}
\usepackage{float}

\begin{document}
\title{Coherent control of subradiant excitations in atomic rings}
\author{Milena Djatchkova}
\affiliation{Institut f\"ur Theoretische Physik, Universit\"at Tübingen, Auf der Morgenstelle 14, 72076 T\"ubingen, Germany}

\author{Igor Lesanovsky}
\affiliation{Institut f\"ur Theoretische Physik, Universit\"at Tübingen, Auf der Morgenstelle 14, 72076 T\"ubingen, Germany}
\affiliation{School of Physics and Astronomy and Centre for the Mathematics and Theoretical Physics of Quantum Non-Equilibrium Systems, The University of Nottingham, Nottingham, NG7 2RD, United Kingdom}

\author{Beatriz Olmos}
\affiliation{Institut f\"ur Theoretische Physik, Universit\"at Tübingen, Auf der Morgenstelle 14, 72076 T\"ubingen, Germany}

\begin{abstract}
   Subradiant excitations in ordered subwavelength atomic arrays are protected against radiative decay, making them a promising platform for storing and manipulating photonic excitations. The external geometry of the array, combined with dynamical control of the atomic dipole orientation, enables localized trapping and coherent transport of these subradiant excitations. Here, we theoretically demonstrate these capabilities in ring-shaped atomic arrays. Specifically, we show adiabatic transport of a localized excitation around a single ring, coherent transfer of a single excitation between two neighboring rings with geometry-controlled selectivity, and interaction-induced conditional phase shifts between two simultaneously trapped excitations in neighboring rings. Depending on the ring geometry, these realize either an entangling iSWAP-type exchange or a controlled-phase operation with tunable conditional phase. Together, these results demonstrate the potential of ordered atomic arrays as a platform for coherent photonic quantum information processing with dissipation-protected collective excitations.
\end{abstract}

\maketitle

\textit{Introduction.---} Collective light-matter interactions arise when quantum emitters couple to a shared electromagnetic environment, allowing photons scattered by different emitters to interfere. This interference gives rise to collective optical modes with radiative properties that differ fundamentally from those of independent emitters, most prominently superradiant and subradiant states whose decay rates are respectively enhanced or suppressed relative to the single-emitter value~\cite{Dicke1954,Lehmberg1970a,Lehmberg1970b,James1993,Chang2018}. First observed in trapped ion pairs~\cite{DeVoe1996}, super- and subradiant emission have since been demonstrated across a wide range of platforms, including cavity-QED settings~\cite{Hotter2023,Pineiro2022,Shankar2021, Kim2018}, molecular systems~\cite{Trebbia2022,McGuyer2014, Mlynek2014}, cold atomic gases~\cite{Ferioli2021,Glicenstein2022}, and geometrically ordered subwavelength arrays,  where the collective optical response reveals signatures of many-body cooperative decay~\cite{Rui2020,Srakaew2023,Douglas2026}. Subradiant states, with their naturally suppressed coupling to the radiation field, have attracted considerable interest as a resource for quantum technologies, including quantum memories, photonic interfaces, and low-loss excitation transport ~\cite{Olmos2013,AsenjoGarcia2017,Reitz2022,Jen2016,Jen2017,Needham2019}.

\begin{figure}
    \centering
    \includegraphics{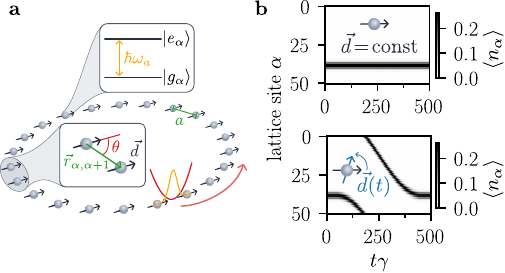}
    \caption{\textbf{Ring setup and subradiant excitation transport.} \textbf{a:} Ring-shaped atomic array with nearest neighbor distance $a$ and parallel in-plane transition dipoles $\vec d$, enclosing an angle $\theta$ with the local bond direction $\vec r_{\alpha,\alpha+1}$. The orientation-dependent interactions create a trapping potential (red) for a localized subradiant excitation (orange), which a rotation of the dipoles coherently transports around the ring. \textbf{b:} Time evolution of the localized excitation for fixed dipoles (top) and during a full adiabatic $2\pi$ dipole rotation over $\tau=400\,\gamma^{-1}$ (bottom), for $N=50$ and $a=0.08\: \lambda_{\mathrm a}$. In both cases the excitation remains localized and loses less than $10^{-5}$ of its population over $t=500\,\gamma^{-1}$.}
    \label{fig1}
\end{figure}
Among these systems, ordered subwavelength atomic arrays provide an especially versatile platform because the electromagnetic fields emitted by neighboring atoms overlap strongly, generating long-range dipole-mediated interactions that govern the collective optical response. As a result, the properties of the collective modes become closely tied to the geometry of the array, allowing them to be engineered through the arrangement of the emitters~\cite{AsenjoGarcia2017,Bettles2016,Bettles2017,Perczel2017,BrowaeysLahaye2020}. Such arrays support exceptionally long-lived subradiant states whose lifetimes can increase exponentially with the number of emitters, while their optical response can be tuned through the orientation of the atomic transition dipoles~\cite{MorenoCardoner2019,Cremer2020,Jen2018a,Jen2018b,Scheil2023,Holzinger2024,Guimond2019}. Protocols for subradiant-protected excitation transport, storage and release of a single photon have recently been investigated for one-dimensional chains \cite{Needham2019}, two-dimensional lattices~\cite{Manzoni2018,Ballantine2021,Ballantine2022,RubiesBigorda2022}, and more recently in ring geometries, where the symmetry of the array gives rise to effective trapping potentials for localized collective excitations~\cite{Cech2023}. The spectral properties of single- and two-excitation subradiant states, including two-photon bound states arising from the saturation of two-level emitters, have likewise been characterized in detail~\cite{AsenjoGarcia2017,Poshakinskiy2021}. Coherent manipulation of such stored excitations, however, remains largely unexplored, although it is essential if they are to become active resources for quantum information processing.

In this work, we show that localized excitations in atomic ring arrays can be not only stored but coherently operated upon while preserving their subradiant protection by using the array geometry and a single global control parameter, the orientation of the transition dipoles. We show both transport of a trapped excitation around a single ring and geometry-selective transfer between two neighboring rings. For two stored excitations, the saturation of the individual emitters renders the excitations interacting. Here, two different geometries realize complementary entangling operations: an iSWAP-type exchange and a controlled-phase operation with tunable conditional phase. These results establish ordered subwavelength arrays as a platform for not only storing but also processing photonic excitations under dissipation protection.

\textit{Model and single-excitation coherent control.---} We consider an array of $N$ identical two-level emitters with ground state $\ket{g_\alpha}$, excited state $\ket{e_\alpha}$, and transition frequency $\omega_{\mathrm a}=2\pi c/\lambda_{\mathrm a}$, as depicted in Fig.~\ref{fig1}a. The emitters interact through the vacuum electromagnetic field and are described by the density matrix $\rho$. Within the Born-Markov approximation, the dynamics obey the master equation
\begin{equation}
    \dot{\rho}=-\frac{\mathrm{i}}{\hbar}\left[H,\rho\right]+{\cal D}(\rho). \label{Lindblad-master}
\end{equation}
The coherent part is governed by the dipole-exchange Hamiltonian
\begin{equation}\label{eq:Ham}
    H=\hbar\sum_{\alpha\neq\beta} V_{\alpha\beta}\sigma_\alpha^\dagger\sigma_\beta,
\end{equation}
where $\sigma_\alpha=\ket{g_\alpha}\!\bra{e_\alpha}$. The exchange couplings \(V_{\alpha\beta}\) depend on both the interatomic separation \(r_{\alpha\beta}\) and the relative orientation of the transition dipoles $\theta_{\alpha\beta}$. Defining \(\kappa_{\alpha\beta}=2\pi r_{\alpha\beta}/\lambda_{\mathrm a}\), they read
\begin{align}
    V_{\alpha \beta}  \!=\! \frac{3\gamma}{4}\Bigg[ \!
    \left( \frac{\cos {\kappa_{\alpha\beta}}}{\kappa_{\alpha\beta}^3} + \frac{\sin {\kappa_{\alpha\beta}}}{\kappa_{\alpha\beta}^2} \right)\!
    g_{\alpha\beta}\!-\!\frac{\cos{\kappa_{\alpha\beta}}}{\kappa_{\alpha\beta}} f_{\alpha\beta} \Bigg] ,
\end{align}
where $\gamma$ is the single-atom decay rate, $f_{\alpha\beta}=\sin^2 \theta_{\alpha\beta}$ and $g_{\alpha\beta}=1-3\cos^2 \theta_{\alpha\beta}$ \cite{AsenjoGarcia2017a}.
Dissipation is described by
\begin{equation}
    {\cal D}(\rho)=\sum_{\alpha \beta} \Gamma_{\alpha \beta} \left( \sigma_\beta \rho \sigma_\alpha^\dagger - \frac{1}{2} \{ \sigma_\alpha^\dagger \sigma_\beta, \rho \}\right),
\end{equation}
with collective dissipation matrix \cite{Jones2018, AsenjoGarcia2017a}
\begin{align}
    \Gamma_{\alpha \beta} \!=\! \frac{3\gamma}{2} \Bigg[ \frac{\sin{\kappa_{\alpha\beta}}}{\kappa_{\alpha\beta}} f_{\alpha\beta}
    \!+ \!\left( \frac{\cos {\kappa_{\alpha\beta}}}{\kappa_{\alpha\beta}^2}-\frac{\sin {\kappa_{\alpha\beta}}}{\kappa_{\alpha\beta}^3}   \right)\!
    g_{\alpha\beta} \Bigg].
\end{align}
For subwavelength emitter separations, photons scattered by neighboring atoms interfere strongly, making the off-diagonal exchange and dissipative couplings significant. The resulting collective radiative modes exhibit enhanced (superradiant) or suppressed (subradiant) decay rates~\cite{Lehmberg1970a}.

We now specialize to the ring geometry shown in Fig.~\ref{fig1}a, consisting of $N$ emitters separated by the nearest neighbor Euclidean distance $a$ and possessing parallel in-plane transition dipole moments. Unless otherwise stated, we use $a=0.08\,\lambda_\mathrm{a}$. However, all results persist over a wider range of subwavelength spacings (see Supplemental Material~\cite{SupMat}).
Because the angle $\theta$ between the dipoles and the local bond direction varies continuously around the ring, the nearest-neighbor exchange coupling becomes position dependent. As shown in Ref.~\cite{Cech2023}, this gives rise to an effective trapping potential whose minimum occurs where the dipoles align with the local bond direction ($\theta=0$). Around this point, the potential is approximately harmonic and supports localized subradiant collective modes. As illustrated in Fig.~\ref{fig1}a, in order to exploit this feature, we initialize the system in a Gaussian wave packet localized near the minimum of the effective trapping potential
\begin{equation}\label{eq:one_exc}
|\psi^\mathrm{1e}_0\rangle = \mathcal{N}\sum_{\alpha} e^{-i k_s (x_\alpha -x_0)} e^{-(x_\alpha-x_0)^2 \sigma^2} \ket{e_\alpha},
\end{equation}
where $x_\alpha=\alpha a$ denotes the position along the ring, $x_0$ the center of the wave packet, $\sigma$ its width in wave-number space, and $\mathcal{N}$ a normalization constant. Choosing the central wave number $k_s$, corresponding to the momentum $\hbar k_s$, outside the light cone, $k_\mathrm{a}<|k_s|<2\pi/a-k_\mathrm{a}$ with $k_\mathrm{a}=2\pi/\lambda_\mathrm{a}$~\cite{AsenjoGarcia2017,Needham2019}, together with a sufficiently narrow wave-number distribution, ensures that the excitation occupies predominantly subradiant collective modes and therefore exhibits exponentially suppressed radiative losses~\cite{Cech2023}. Within the single-excitation manifold, the dynamics in the absence of photon emission is generated by the effective non-Hermitian Hamiltonian~\cite{Needham2019}
\begin{equation}\label{eq:Heff}
    H^\mathrm{eff}_{\alpha \beta} = \hbar \left(V_{\alpha \beta} - \frac{\mathrm{i}}{2} \Gamma_{\alpha \beta}\right).
\end{equation}
As shown in the upper panel of Fig.~\ref{fig1}b, the excitation remains as well localized near the trapping minimum for times far exceeding the lifetime of an individual emitter.

\begin{figure*}
    \centering
    \includegraphics{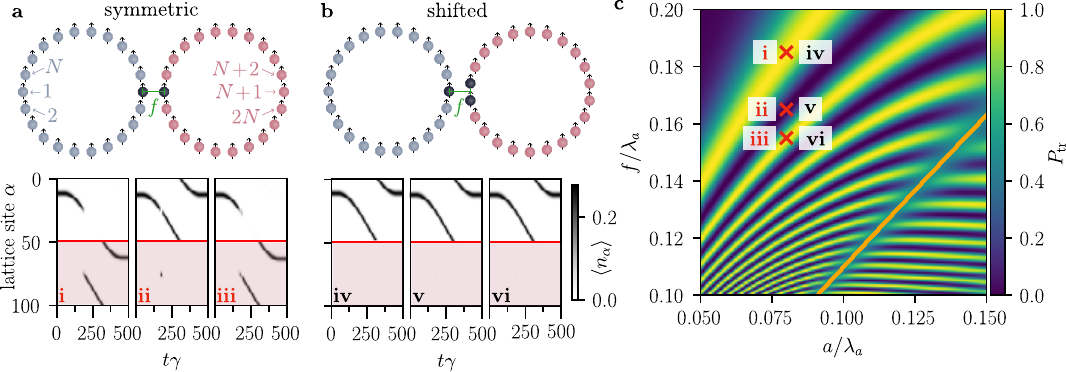}
    \caption{\textbf{Controlled excitation exchange dynamics between coupled rings.} \textbf{a:} Symmetric configuration with axially aligned rings. \textbf{b:} Shifted configuration, where one ring is rotated such that the atoms at the point of closest proximity form a triangle. In both cases, the inter-ring separation is denoted as $f$.
    \textbf{c:} Excitation transfer probability in the symmetric configuration for $N=50$, shown as a function of interatomic distance $a$ and ring separation $f$. The solid orange line indicates an energy crossing between the participating eigenstates. \textbf{i-iii:} Representative time evolution of the site occupations for $a=0.08\:\lambda_a$, illustrating one ($f=0.185\:\lambda_a$), two ($f=0.165\:\lambda_a$) and three ($f=0.155\:\lambda_a$) coherent oscillation half-cycles between the rings. Red crosses mark the corresponding parameter sets in panel \textbf{c}. \textbf{iv-vi}: Time evolution for the shifted configuration using the same values of $a$ and $f$, showing the suppression of the excitation transfer.
    }
    \label{fig2}
\end{figure*}

Since the trapping potential is determined by the dipole orientation, a collective in-plane rotation of the dipoles translates the potential minimum around the ring, as illustrated in Fig.~\ref{fig1}a (see below for a possible implementation). Provided the rotation is sufficiently slow, the localized excitation adiabatically follows the moving potential. More precisely, adiabatic evolution requires $\tau\gg\hbar/|\Delta E|$, where $\Delta E$ denotes the energy gap between the trapped mode and the nearest collective eigenmode of the effective Hamiltonian. As an example, we perform a complete $2\pi$ rotation over a duration $\tau=400\,\gamma^{-1}$ according to
\begin{equation}\label{eq:dipole}
\vec{d}(t)\!=\!\left(\!\!\cos\left[2\pi \sin^2\!\left(\frac{\pi t}{2\tau}\right)\right]\! ;\sin\left[2\pi \sin^2\!\left(\frac{\pi t}{2\tau}\right)\right]\!; 0\!\right)\!.
\end{equation}
This smooth profile ensures that the angular velocity vanishes at the beginning and end of the protocol, thereby suppressing non-adiabatic transitions beyond leading order in $1/\tau$. Consequently, the excitation remains localized and subradiant throughout transport despite the continuous manipulation of the trapping potential, as shown in the lower panel of Fig.~\ref{fig1}b.

To quantify the robustness of the protocol, we monitor the collective excitation survival probability,
\begin{equation}
    P_\mathrm{sur}(t)=\sum_{\alpha}\left< n_\alpha(t)\right>,
\end{equation}
where $n_\alpha=\sigma_\alpha^\dagger\sigma_\alpha$. Initially $P_{\mathrm{sur}}(0)=1$, and radiative losses lead to a gradual decrease with time. Remarkably, in all manipulation protocols considered in this work, the survival probability remains close to unity, demonstrating that coherent transport preserves both the localization and the subradiant protection of the collective excitation. Numerical convergence of all reported values is documented in \cite{SupMat}.

\textit{Single-excitation transfer between ring lattices.---} We now consider two identical rings separated by a distance $f$, arranged either in the \textit{symmetric} or \textit{shifted} geometry shown in Fig.~\ref{fig2}a,b. The qualitative difference between them is that only the symmetric geometry supports hybridization of the localized subradiant modes, whereas the shifted geometry preserves their localization.

As before, the excitation is initially localized in the left ring and transported by the adiabatic full dipole rotation of Eq.~\eqref{eq:dipole}. For sufficiently close rings, the excitation is fully or partially
transferred to the second ring, depending on their distance and relative
orientation. To quantify the excitation transfer probability, we define
\begin{equation}
    P_\mathrm{tr}=\sum_{\alpha=N+1}^{2N}\left<n_\alpha(t_\mathrm{end})\right>,
\end{equation}
which measures the excitation population in the second, initially empty ring, at the final time $t_\mathrm{end}$ such that a complete transfer corresponds to $P_\mathrm{tr}= 1$.

In the symmetric configuration, the localized subradiant modes hybridize in the region where the rings are closest, forming delocalized eigenstates extending over both rings. As the transported wave packet enters this region, it overlaps with these hybridized modes and coherently tunnels between the rings. The resulting transfer probability depends on the competition between intra-ring transport and inter-ring coupling, producing the interference pattern shown in Fig.~\ref{fig2}c. The interruption of the fringe pattern is the manifestation of an energy crossing between the participating subradiant modes (see Supplemental Material \cite{SupMat} for details).

The site occupations $\langle n_\alpha\rangle$, shown in Fig.~\ref{fig2}i--iii, reveal the transfer dynamics. As the wave packet enters the hybridization region, it undergoes coherent Rabi-like oscillations between the two rings. Depending on the ring separations, the evolution terminates after either an odd or even half-cycle resulting, respectively, in complete transfer to the second ring ($P_\mathrm{tr}=0.996$ and $0.981$ for panels i and iii) or return to the original ring with an additional phase. Intermediate separations produce coherent partial transfer, corresponding to a controllable splitting of the wave packet between the two rings~\cite{SupMat}. Throughout, the participating collective modes remain strongly subradiant, yielding survival probabilities above $0.95$ and exceeding $0.9999$ before the energy crossing. The transfer is, moreover, robust against misalignment of the rings \cite{SupMat}.

In contrast, the shifted geometry exhibits essentially no hybridization of the localized subradiant modes. Consequently, the excitation remains confined to its original ring despite the small inter-ring separation. Panels iv-vi illustrate this, using the same values of the lattice spacing $a$ and the ring separation $f$ as panels i-iii, respectively, but for the shifted configuration. Despite the very small ring separations, no excitation transfer, complete or partial, is observed: the two rings are effectively decoupled in the shifted configuration.

\begin{figure}
    \centering
    \includegraphics[width=0.47\textwidth]{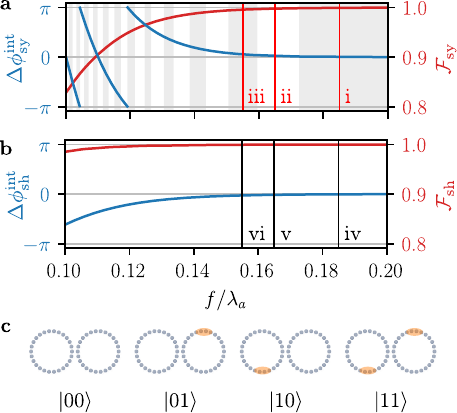}
    \caption{\textbf{Conditional phase accumulation.} \textbf{a,b:} Interaction-dependent part $\Delta\phi^\mathrm{int}$ of the excess phase and fidelity $\mathcal{F}$ as a function of the inter-ring separation $f$ for the symmetric (\textbf{a}) and shifted (\textbf{b}) configurations, with fixed interatomic spacing $a=0.08\,\lambda_\mathrm{a}$. Shaded regions in \textbf{a} indicate $P_\mathrm{tr}>1/2$, where the exchange contribution is $\Delta\phi^\mathrm{ex}=\pi$. Elsewhere, and throughout \textbf{b}, $\Delta\phi^\mathrm{ex}=0$. The vertical lines labeled \textbf{i--vi} indicate the parameter values of the time evolutions shown in Fig.~\ref{fig2}. \textbf{c:} Logical basis defined by the presence or absence of a trapped excitation in each ring.}
    \label{fig3}
\end{figure}

\textit{Coherent interactions between subradiant excitations.---} Having established coherent transport within a ring and selective transfer between neighboring rings, we finally show that two trapped subradiant excitations can interact coherently while preserving their subradiant protection. To this end, we initialize one localized excitation in each ring and simultaneously steer the two wave packets by adiabatically rotating the dipole moments. As the excitations approach each other, they interact through the saturation of the individual emitters, which cannot be doubly excited, giving rise to an interaction-induced phase. This phase is extracted by comparing the evolution with that of two independent excitations under the same protocol.

The dynamics of two excitations is confined to $\mathcal{H}_{\leq 2} = \mathcal{H}_0 \oplus \mathcal{H}_1 \oplus \mathcal{H}_2$, spanned by the vacuum $\ket{G}=\ket{g_1\dots g_{2N}}$, the $2N$ single-excitation states $\ket{e_\alpha}=\ket{g_1\dots e_\alpha\dots g_{2N}}$, and the $\binom{2N}{2}$ two-excitation states $\ket{e_\alpha e_\beta} \!=\! \ket{g_1 \dots e_\alpha \dots e_\beta \dots g_{2N}}$. Because the coherent Hamiltonian conserves the excitation number and no external drive repopulates the two-excitation sector, each radiative quantum jump transfers population irreversibly from $\mathcal{H}_2$ to $\mathcal{H}_1$. As detailed in the Supplemental Material~\cite{SupMat}, the two-excitation block therefore evolves exactly under $H^\mathrm{eff}$, and trajectories containing an emission event do not contribute to the quantities considered below. Correspondingly, the Liouvillian spectrum in this sector follows from that of $H^\mathrm{eff}$~\cite{Leppenen2026}.

As initial state, we choose the symmetrized product $|\psi_0^{2e}\rangle$ of two wave packets of the form~\eqref{eq:one_exc}, one in each ring, centered at $x_{0,1}$ and $x_{0,2}$ and normalized within $\mathcal{H}_2$. The preservation of this state is quantified by the fidelity ${\cal F}=|\langle\psi_0^\mathrm{2e}|\psi(t_\mathrm{end})\rangle|^2$, which exceeds 0.98 for $f\gtrsim 0.13\,\lambda_\mathrm{a}$ in the symmetric configuration and throughout the shown range in the shifted one (Fig.~\ref{fig3}a,b), showing that at the end of the protocol the excitations predominantly return to the intended subradiant initial wave packet up to a phase. Although the exchange Hamiltonian~\eqref{eq:Ham} is quadratic, the two excitations are not independent: a two-level emitter cannot be doubly excited, the same saturation nonlinearity that underlies subradiant two-photon bound states~\cite{Poshakinskiy2021}. Without this constraint, the excitations would propagate independently, yielding the state $|\psi_\mathrm{ind}(t)\rangle$. We therefore define the interaction-induced excess phase as
\begin{equation}\label{eq:phase}
  \Delta\phi=\arg\frac{\langle\psi_0^{2e}|\psi(t_\mathrm{end})\rangle}
                      {\langle\psi_0^{2e}|\psi_\mathrm{ind}(t_\mathrm{end})\rangle},
\end{equation}
in which all single-excitation phases, including those acquired during transfer, cancel so that $\Delta\phi$ vanishes identically for
non-interacting excitations.

In the symmetric configuration, each component of a single excitation that is transferred to the other ring acquires a phase $\pi/2$~\cite{SupMat}. For two independent excitations, which are transferred simultaneously in opposite directions, the reference amplitude in the denominator therefore changes sign at the ring separations where $P_\mathrm{tr}=1/2$, similar to Hong--Ou--Mandel interference. We absorb these jumps into an exchange contribution, $\Delta\phi_\mathrm{sy}=\Delta\phi^\mathrm{ex}_\mathrm{sy}+\Delta\phi^\mathrm{int}_\mathrm{sy}$, with $\Delta\phi^\mathrm{ex}_\mathrm{sy}=\pi$ where the excitations are predominantly exchanged ($P_\mathrm{tr}>1/2$) and $\Delta\phi^\mathrm{ex}_\mathrm{sy}=0$ otherwise. Figure~\ref{fig3}a shows the remaining interaction-dependent part $\Delta\phi^\mathrm{int}_\mathrm{sy}$, with $\Delta\phi^\mathrm{ex}_\mathrm{sy}$ indicated by shading. In the shifted configuration, no transfer occurs, so that $\Delta\phi^\mathrm{ex}_\mathrm{sh}=0$, and $\Delta\phi^\mathrm{int}_\mathrm{sh}$ grows smoothly as the rings approach (Fig.~\ref{fig3}b). In the symmetric configuration, the additional interaction-induced phase at small separations comes with additional radiative loss, which reduces $\mathcal{F}_\mathrm{sy}$.

Experimentally, $\Delta\phi$ can be accessed by a conditional Ramsey measurement: one ring is prepared in a superposition of zero and one excitation, e.g., by a $\pi/2$-pulse in the preparation scheme of Ref.~\cite{Cech2023}, the other is either empty or occupied, and the phase of the first excitation after its release is compared between the two cases. Conditional phases of this kind have been measured interferometrically for Rydberg-mediated photon--photon interactions~\cite{Tiarks2016}.

Beyond characterizing coherent interactions between protected collective excitations, these results also admit a natural interpretation in terms of elementary photonic quantum operations. Encoding logical states in the presence or absence of a trapped excitation in each ring, as illustrated in Fig.~\ref{fig3}c, the two geometries realize complementary elementary transformations.

The symmetric geometry extends the single-excitation transfer dynamics discussed in the previous section to the two-excitation manifold. The one-excitation states $\ket{10}$ and $\ket{01}$ continue to undergo coherent mode mixing, with the effective splitting ratio controlled by the ring separation. This evolution is formally equivalent to a beam-splitter operation acting on the stored subradiant excitations (see Supplemental Material~\cite{SupMat} for a quantitative analysis). At full transfer, the protocol maps $|10\rangle\to \mathrm{i}|01\rangle$ and $|01\rangle\to \mathrm{i}|10\rangle$, while the constrained evolution leaves $|11\rangle\to|11\rangle$. Up to single-ring phases, this is the iSWAP operation generated by an exchange coupling between two-level systems, which is entangling, whereas independent excitations would acquire a phase $\pi$ on $|11\rangle$, amounting to a mere swap. Since $\Delta\phi^\mathrm{int}_\mathrm{sy}$ remains small for $f\geq0.15\,\lambda_\mathrm{a}$, the conditional phase of this operation is set by the constraint rather than by the precise ring separation.

The shifted geometry exhibits complementary behavior. Here, excitation transfer remains strongly suppressed, so that the logical basis states are preserved throughout the protocol. Apart from single-ring phases, which can be removed by local operations, the only nontrivial evolution is the interaction-induced phase of the doubly occupied state, $|11\rangle\to e^{i\Delta\phi^\mathrm{int}_\mathrm{sh}}|11\rangle$. In the limit of negligible radiative losses, the protocol thus realizes a controlled-phase operation whose conditional phase is tuned continuously by the ring separation, reaching a controlled-Z gate, $|\Delta\phi^\mathrm{int}_\mathrm{sh}|=\pi$, at $f\approx0.09\,\lambda_\mathrm{a}$ with $\mathcal{F}_\mathrm{sh}=0.96$, beyond the range shown in Fig.~\ref{fig3}b.

Together with the coherent transport and transfer protocols demonstrated above, these results establish a versatile toolbox for the coherent manipulation of protected collective excitations in atomic ring arrays.

\textit{Experimental implementation.} --- The protocols rely on three ingredients. First, the emitters must possess a closed two-level transition with a linear transition dipole, since decay into other states is not collectively suppressed. This holds, for instance, for $J=0\rightarrow J=1$ transitions without branching, and for artificial emitters such as superconducting qubits~\cite{Du2026}. Second, the emitters must be arranged in rings at subwavelength spacings. Optical tweezers allow arbitrary two-dimensional arrangements~\cite{Rui2020,Srakaew2023,Douglas2026,BrowaeysLahaye2020}, and long-wavelength transitions relax the absolute length scales: for the mid-infrared transitions of alkaline-earth atoms~\cite{Olmos2013,Holman2026}, with $\lambda_\mathrm{a}\approx3\,\mu$m, $a=0.2\,\lambda_\mathrm{a}$ corresponds to about 600\,nm, where all protocols remain operational~\cite{SupMat}. Third, the dipole orientation must be controlled globally. For the $\pi$ component of a Zeeman-resolved transition, the dipole is parallel to a homogeneous in-plane bias field, so that rotating this field rotates all dipoles identically and simultaneously, provided the Zeeman splitting exceeds the dipolar couplings and the rotation is slow compared with it \cite{Needham2019}. The localized initial state can be prepared, and later released, with the Rydberg-blockade scheme of Ref.~\cite{Cech2023}, which creates exactly one excitation and imprints a central wave number outside the light cone.

\textit{Conclusion and outlook.---} We have shown that localized subradiant excitations in atomic ring arrays can be coherently manipulated while preserving their protection against radiative decay using a single global control parameter, the dipole orientation. Moreover, we have demonstrated that these protocols naturally admit an interpretation in terms of elementary photonic quantum operations: depending on the relative geometry of neighboring rings, the dynamics may realize either an exchange operation, which at full transfer becomes an entangling iSWAP, or a controlled-phase operation with a conditional phase tunable by the ring separation. %

Looking ahead, extending these ideas to larger networks of coupled rings could enable programmable routing, interference, and processing of protected collective excitations. In this direction, it will be key to further improve the fidelities and transfer probabilities reported here, for example by using optimal control. An important direction will be to investigate the robustness of these protocols against disorder and atomic motion~\cite{Olmos2025,Rubies2025}, as well as their realization in the platforms discussed above.

\textit{Acknowledgments.---} The authors acknowledge support by the state of Baden-Württemberg through bwHPC and the German Research Foundation (DFG) through grant no INST 40/575-1 FUGG (JUSTUS 2 cluster). This work is supported by the ERC grant OPEN-2QS (Grant No. 101164443).  We acknowledge funding from the Deutsche Forschungsgemeinschaft (DFG, German Research Foundation) through the Research Unit FOR 5413/1, Grant No. 465199066, and through JST-DFG 2024: Japanese-German Joint Call for Proposals on “Quantum Technologies” (Japan-JST-DFG-ASPIRE 2024) under DFG Grant No. 55456179.

\bibliography{references.bib}

\begin{thebibliography}{50}%
\makeatletter
\providecommand \@ifxundefined [1]{%
 \@ifx{#1\undefined}
}%
\providecommand \@ifnum [1]{%
 \ifnum #1\expandafter \@firstoftwo
 \else \expandafter \@secondoftwo
 \fi
}%
\providecommand \@ifx [1]{%
 \ifx #1\expandafter \@firstoftwo
 \else \expandafter \@secondoftwo
 \fi
}%
\providecommand \natexlab [1]{#1}%
\providecommand \enquote  [1]{``#1''}%
\providecommand \bibnamefont  [1]{#1}%
\providecommand \bibfnamefont [1]{#1}%
\providecommand \citenamefont [1]{#1}%
\providecommand \href@noop [0]{\@secondoftwo}%
\providecommand \href [0]{\begingroup \@sanitize@url \@href}%
\providecommand \@href[1]{\@@startlink{#1}\@@href}%
\providecommand \@@href[1]{\endgroup#1\@@endlink}%
\providecommand \@sanitize@url [0]{\catcode `\\12\catcode `\$12\catcode
  `\&12\catcode `\#12\catcode `\^12\catcode `\_12\catcode `\%12\relax}%
\providecommand \@@startlink[1]{}%
\providecommand \@@endlink[0]{}%
\providecommand \url  [0]{\begingroup\@sanitize@url \@url }%
\providecommand \@url [1]{\endgroup\@href {#1}{\urlprefix }}%
\providecommand \urlprefix  [0]{URL }%
\providecommand \Eprint [0]{\href }%
\providecommand \doibase [0]{https://doi.org/}%
\providecommand \selectlanguage [0]{\@gobble}%
\providecommand \bibinfo  [0]{\@secondoftwo}%
\providecommand \bibfield  [0]{\@secondoftwo}%
\providecommand \translation [1]{[#1]}%
\providecommand \BibitemOpen [0]{}%
\providecommand \bibitemStop [0]{}%
\providecommand \bibitemNoStop [0]{.\EOS\space}%
\providecommand \EOS [0]{\spacefactor3000\relax}%
\providecommand \BibitemShut  [1]{\csname bibitem#1\endcsname}%
\let\auto@bib@innerbib\@empty
\bibitem [{\citenamefont {Dicke}(1954)}]{Dicke1954}%
  \BibitemOpen
  \bibfield  {author} {\bibinfo {author} {\bibfnamefont {R.~H.}\ \bibnamefont
  {Dicke}},\ }\bibfield  {title} {\bibinfo {title} {Coherence in spontaneous
  radiation processes},\ }\href {https://doi.org/10.1103/PhysRev.93.99}
  {\bibfield  {journal} {\bibinfo  {journal} {Phys. Rev.}\ }\textbf {\bibinfo
  {volume} {93}},\ \bibinfo {pages} {99} (\bibinfo {year} {1954})}\BibitemShut
  {NoStop}%
\bibitem [{\citenamefont {Lehmberg}(1970{\natexlab{a}})}]{Lehmberg1970a}%
  \BibitemOpen
  \bibfield  {author} {\bibinfo {author} {\bibfnamefont {R.~H.}\ \bibnamefont
  {Lehmberg}},\ }\bibfield  {title} {\bibinfo {title} {Radiation from an
  $n$-atom system. {I}. general formalism},\ }\href
  {https://doi.org/10.1103/PhysRevA.2.883} {\bibfield  {journal} {\bibinfo
  {journal} {Phys. Rev. A}\ }\textbf {\bibinfo {volume} {2}},\ \bibinfo {pages}
  {883} (\bibinfo {year} {1970}{\natexlab{a}})}\BibitemShut {NoStop}%
\bibitem [{\citenamefont {Lehmberg}(1970{\natexlab{b}})}]{Lehmberg1970b}%
  \BibitemOpen
  \bibfield  {author} {\bibinfo {author} {\bibfnamefont {R.~H.}\ \bibnamefont
  {Lehmberg}},\ }\bibfield  {title} {\bibinfo {title} {Radiation from an
  $n$-atom system. {II}. spontaneous emission from a pair of atoms},\ }\href
  {https://doi.org/10.1103/PhysRevA.2.889} {\bibfield  {journal} {\bibinfo
  {journal} {Phys. Rev. A}\ }\textbf {\bibinfo {volume} {2}},\ \bibinfo {pages}
  {889} (\bibinfo {year} {1970}{\natexlab{b}})}\BibitemShut {NoStop}%
\bibitem [{\citenamefont {James}(1993)}]{James1993}%
  \BibitemOpen
  \bibfield  {author} {\bibinfo {author} {\bibfnamefont {D.~F.~V.}\
  \bibnamefont {James}},\ }\bibfield  {title} {\bibinfo {title} {Frequency
  shifts in spontaneous emission from two interacting atoms},\ }\href
  {https://doi.org/10.1103/PhysRevA.47.1336} {\bibfield  {journal} {\bibinfo
  {journal} {Phys. Rev. A}\ }\textbf {\bibinfo {volume} {47}},\ \bibinfo
  {pages} {1336} (\bibinfo {year} {1993})}\BibitemShut {NoStop}%
\bibitem [{\citenamefont {Chang}\ \emph {et~al.}(2018)\citenamefont {Chang},
  \citenamefont {Douglas}, \citenamefont {Gonz{\'a}lez-Tudela}, \citenamefont
  {Hung},\ and\ \citenamefont {Kimble}}]{Chang2018}%
  \BibitemOpen
  \bibfield  {author} {\bibinfo {author} {\bibfnamefont {D.~E.}\ \bibnamefont
  {Chang}}, \bibinfo {author} {\bibfnamefont {J.~S.}\ \bibnamefont {Douglas}},
  \bibinfo {author} {\bibfnamefont {A.}~\bibnamefont {Gonz{\'a}lez-Tudela}},
  \bibinfo {author} {\bibfnamefont {C.-L.}\ \bibnamefont {Hung}},\ and\
  \bibinfo {author} {\bibfnamefont {H.~J.}\ \bibnamefont {Kimble}},\ }\bibfield
   {title} {\bibinfo {title} {Colloquium: Quantum matter built from nanoscopic
  lattices of atoms and photons},\ }\href
  {https://doi.org/10.1103/RevModPhys.90.031002} {\bibfield  {journal}
  {\bibinfo  {journal} {Rev. Mod. Phys.}\ }\textbf {\bibinfo {volume} {90}},\
  \bibinfo {pages} {031002} (\bibinfo {year} {2018})}\BibitemShut {NoStop}%
\bibitem [{\citenamefont {DeVoe}\ and\ \citenamefont
  {Brewer}(1996)}]{DeVoe1996}%
  \BibitemOpen
  \bibfield  {author} {\bibinfo {author} {\bibfnamefont {R.~G.}\ \bibnamefont
  {DeVoe}}\ and\ \bibinfo {author} {\bibfnamefont {R.~G.}\ \bibnamefont
  {Brewer}},\ }\bibfield  {title} {\bibinfo {title} {Observation of
  superradiant and subradiant spontaneous emission of two trapped ions},\
  }\href {https://doi.org/10.1103/PhysRevLett.76.2049} {\bibfield  {journal}
  {\bibinfo  {journal} {Phys. Rev. Lett.}\ }\textbf {\bibinfo {volume} {76}},\
  \bibinfo {pages} {2049} (\bibinfo {year} {1996})}\BibitemShut {NoStop}%
\bibitem [{\citenamefont {Hotter}\ \emph {et~al.}(2023)\citenamefont {Hotter},
  \citenamefont {Ostermann},\ and\ \citenamefont {Ritsch}}]{Hotter2023}%
  \BibitemOpen
  \bibfield  {author} {\bibinfo {author} {\bibfnamefont {C.}~\bibnamefont
  {Hotter}}, \bibinfo {author} {\bibfnamefont {L.}~\bibnamefont {Ostermann}},\
  and\ \bibinfo {author} {\bibfnamefont {H.}~\bibnamefont {Ritsch}},\
  }\bibfield  {title} {\bibinfo {title} {Cavity sub- and superradiance for
  transversely driven atomic ensembles},\ }\href
  {https://doi.org/10.1103/PhysRevResearch.5.013056} {\bibfield  {journal}
  {\bibinfo  {journal} {Phys. Rev. Res.}\ }\textbf {\bibinfo {volume} {5}},\
  \bibinfo {pages} {013056} (\bibinfo {year} {2023})}\BibitemShut {NoStop}%
\bibitem [{\citenamefont {Pi{\~n}eiro~Orioli}\ \emph
  {et~al.}(2022)\citenamefont {Pi{\~n}eiro~Orioli}, \citenamefont {Thompson},\
  and\ \citenamefont {Rey}}]{Pineiro2022}%
  \BibitemOpen
  \bibfield  {author} {\bibinfo {author} {\bibfnamefont {A.}~\bibnamefont
  {Pi{\~n}eiro~Orioli}}, \bibinfo {author} {\bibfnamefont {J.~K.}\ \bibnamefont
  {Thompson}},\ and\ \bibinfo {author} {\bibfnamefont {A.~M.}\ \bibnamefont
  {Rey}},\ }\bibfield  {title} {\bibinfo {title} {Emergent dark states from
  superradiant dynamics in multilevel atoms in a cavity},\ }\href
  {https://doi.org/10.1103/PhysRevX.12.011054} {\bibfield  {journal} {\bibinfo
  {journal} {Phys. Rev. X}\ }\textbf {\bibinfo {volume} {12}},\ \bibinfo
  {pages} {011054} (\bibinfo {year} {2022})}\BibitemShut {NoStop}%
\bibitem [{\citenamefont {Shankar}\ \emph {et~al.}(2021)\citenamefont
  {Shankar}, \citenamefont {Reilly}, \citenamefont {J{\"a}ger},\ and\
  \citenamefont {Holland}}]{Shankar2021}%
  \BibitemOpen
  \bibfield  {author} {\bibinfo {author} {\bibfnamefont {A.}~\bibnamefont
  {Shankar}}, \bibinfo {author} {\bibfnamefont {J.~T.}\ \bibnamefont {Reilly}},
  \bibinfo {author} {\bibfnamefont {S.~B.}\ \bibnamefont {J{\"a}ger}},\ and\
  \bibinfo {author} {\bibfnamefont {M.~J.}\ \bibnamefont {Holland}},\
  }\bibfield  {title} {\bibinfo {title} {Subradiant-to-subradiant phase
  transition in the bad cavity laser},\ }\href
  {https://doi.org/10.1103/PhysRevLett.127.073603} {\bibfield  {journal}
  {\bibinfo  {journal} {Phys. Rev. Lett.}\ }\textbf {\bibinfo {volume} {127}},\
  \bibinfo {pages} {073603} (\bibinfo {year} {2021})}\BibitemShut {NoStop}%
\bibitem [{\citenamefont {Kim}\ \emph {et~al.}(2018)\citenamefont {Kim},
  \citenamefont {Yang}, \citenamefont {Oh},\ and\ \citenamefont
  {An}}]{Kim2018}%
  \BibitemOpen
  \bibfield  {author} {\bibinfo {author} {\bibfnamefont {J.}~\bibnamefont
  {Kim}}, \bibinfo {author} {\bibfnamefont {D.}~\bibnamefont {Yang}}, \bibinfo
  {author} {\bibfnamefont {S.-h.}\ \bibnamefont {Oh}},\ and\ \bibinfo {author}
  {\bibfnamefont {K.}~\bibnamefont {An}},\ }\bibfield  {title} {\bibinfo
  {title} {Coherent single-atom superradiance},\ }\href
  {https://doi.org/10.1126/science.aar2179} {\bibfield  {journal} {\bibinfo
  {journal} {Science}\ }\textbf {\bibinfo {volume} {359}},\ \bibinfo {pages}
  {662} (\bibinfo {year} {2018})}\BibitemShut {NoStop}%
\bibitem [{\citenamefont {Trebbia}\ \emph {et~al.}(2022)\citenamefont
  {Trebbia}, \citenamefont {Deplano}, \citenamefont {Tamarat},\ and\
  \citenamefont {Lounis}}]{Trebbia2022}%
  \BibitemOpen
  \bibfield  {author} {\bibinfo {author} {\bibfnamefont {J.-B.}\ \bibnamefont
  {Trebbia}}, \bibinfo {author} {\bibfnamefont {Q.}~\bibnamefont {Deplano}},
  \bibinfo {author} {\bibfnamefont {P.}~\bibnamefont {Tamarat}},\ and\ \bibinfo
  {author} {\bibfnamefont {B.}~\bibnamefont {Lounis}},\ }\bibfield  {title}
  {\bibinfo {title} {Tailoring the superradiant and subradiant nature of two
  coherently coupled quantum emitters},\ }\href
  {https://doi.org/10.1038/s41467-022-30672-2} {\bibfield  {journal} {\bibinfo
  {journal} {Nat. Commun.}\ }\textbf {\bibinfo {volume} {13}},\ \bibinfo
  {pages} {2962} (\bibinfo {year} {2022})}\BibitemShut {NoStop}%
\bibitem [{\citenamefont {McGuyer}\ \emph {et~al.}(2014)\citenamefont
  {McGuyer}, \citenamefont {McDonald}, \citenamefont {Iwata}, \citenamefont
  {Tarallo}, \citenamefont {Skomorowski}, \citenamefont {Moszynski},\ and\
  \citenamefont {Zelevinsky}}]{McGuyer2014}%
  \BibitemOpen
  \bibfield  {author} {\bibinfo {author} {\bibfnamefont {B.~H.}\ \bibnamefont
  {McGuyer}}, \bibinfo {author} {\bibfnamefont {M.}~\bibnamefont {McDonald}},
  \bibinfo {author} {\bibfnamefont {G.~Z.}\ \bibnamefont {Iwata}}, \bibinfo
  {author} {\bibfnamefont {M.~G.}\ \bibnamefont {Tarallo}}, \bibinfo {author}
  {\bibfnamefont {W.}~\bibnamefont {Skomorowski}}, \bibinfo {author}
  {\bibfnamefont {R.}~\bibnamefont {Moszynski}},\ and\ \bibinfo {author}
  {\bibfnamefont {T.}~\bibnamefont {Zelevinsky}},\ }\bibfield  {title}
  {\bibinfo {title} {Precise study of asymptotic physics with subradiant
  ultracold molecules},\ }\href {https://doi.org/10.1038/nphys3182} {\bibfield
  {journal} {\bibinfo  {journal} {Nat. Phys.}\ }\textbf {\bibinfo {volume}
  {11}},\ \bibinfo {pages} {32} (\bibinfo {year} {2014})}\BibitemShut {NoStop}%
\bibitem [{\citenamefont {Mlynek}\ \emph {et~al.}(2014)\citenamefont {Mlynek},
  \citenamefont {Abdumalikov}, \citenamefont {Eichler},\ and\ \citenamefont
  {Wallraff}}]{Mlynek2014}%
  \BibitemOpen
  \bibfield  {author} {\bibinfo {author} {\bibfnamefont {J.~A.}\ \bibnamefont
  {Mlynek}}, \bibinfo {author} {\bibfnamefont {A.~A.}\ \bibnamefont
  {Abdumalikov}}, \bibinfo {author} {\bibfnamefont {C.}~\bibnamefont
  {Eichler}},\ and\ \bibinfo {author} {\bibfnamefont {A.}~\bibnamefont
  {Wallraff}},\ }\bibfield  {title} {\bibinfo {title} {Observation of {Dicke}
  superradiance for two artificial atoms in a cavity with high decay rate},\
  }\href {https://doi.org/10.1038/ncomms6186} {\bibfield  {journal} {\bibinfo
  {journal} {Nat. Commun.}\ }\textbf {\bibinfo {volume} {5}},\ \bibinfo {pages}
  {5186} (\bibinfo {year} {2014})}\BibitemShut {NoStop}%
\bibitem [{\citenamefont {Ferioli}\ \emph {et~al.}(2021)\citenamefont
  {Ferioli}, \citenamefont {Glicenstein}, \citenamefont {Henriet},
  \citenamefont {Ferrier-Barbut},\ and\ \citenamefont
  {Browaeys}}]{Ferioli2021}%
  \BibitemOpen
  \bibfield  {author} {\bibinfo {author} {\bibfnamefont {G.}~\bibnamefont
  {Ferioli}}, \bibinfo {author} {\bibfnamefont {A.}~\bibnamefont
  {Glicenstein}}, \bibinfo {author} {\bibfnamefont {L.}~\bibnamefont
  {Henriet}}, \bibinfo {author} {\bibfnamefont {I.}~\bibnamefont
  {Ferrier-Barbut}},\ and\ \bibinfo {author} {\bibfnamefont {A.}~\bibnamefont
  {Browaeys}},\ }\bibfield  {title} {\bibinfo {title} {Storage and release of
  subradiant excitations in a dense atomic cloud},\ }\href
  {https://doi.org/10.1103/PhysRevX.11.021031} {\bibfield  {journal} {\bibinfo
  {journal} {Phys. Rev. X}\ }\textbf {\bibinfo {volume} {11}},\ \bibinfo
  {pages} {021031} (\bibinfo {year} {2021})}\BibitemShut {NoStop}%
\bibitem [{\citenamefont {Glicenstein}\ \emph {et~al.}(2022)\citenamefont
  {Glicenstein}, \citenamefont {Ferioli}, \citenamefont {Browaeys},\ and\
  \citenamefont {Ferrier-Barbut}}]{Glicenstein2022}%
  \BibitemOpen
  \bibfield  {author} {\bibinfo {author} {\bibfnamefont {A.}~\bibnamefont
  {Glicenstein}}, \bibinfo {author} {\bibfnamefont {G.}~\bibnamefont
  {Ferioli}}, \bibinfo {author} {\bibfnamefont {A.}~\bibnamefont {Browaeys}},\
  and\ \bibinfo {author} {\bibfnamefont {I.}~\bibnamefont {Ferrier-Barbut}},\
  }\bibfield  {title} {\bibinfo {title} {From superradiance to subradiance:
  exploring the many-body {Dicke} ladder},\ }\href
  {https://doi.org/10.1364/OL.451903} {\bibfield  {journal} {\bibinfo
  {journal} {Opt. Lett.}\ }\textbf {\bibinfo {volume} {47}},\ \bibinfo {pages}
  {1541} (\bibinfo {year} {2022})}\BibitemShut {NoStop}%
\bibitem [{\citenamefont {Rui}\ \emph {et~al.}(2020)\citenamefont {Rui},
  \citenamefont {Wei}, \citenamefont {Rubio-Abadal}, \citenamefont {Hollerith},
  \citenamefont {Zeiher}, \citenamefont {Stamper-Kurn}, \citenamefont {Gross},\
  and\ \citenamefont {Bloch}}]{Rui2020}%
  \BibitemOpen
  \bibfield  {author} {\bibinfo {author} {\bibfnamefont {J.}~\bibnamefont
  {Rui}}, \bibinfo {author} {\bibfnamefont {D.}~\bibnamefont {Wei}}, \bibinfo
  {author} {\bibfnamefont {A.}~\bibnamefont {Rubio-Abadal}}, \bibinfo {author}
  {\bibfnamefont {S.}~\bibnamefont {Hollerith}}, \bibinfo {author}
  {\bibfnamefont {J.}~\bibnamefont {Zeiher}}, \bibinfo {author} {\bibfnamefont
  {D.~M.}\ \bibnamefont {Stamper-Kurn}}, \bibinfo {author} {\bibfnamefont
  {C.}~\bibnamefont {Gross}},\ and\ \bibinfo {author} {\bibfnamefont
  {I.}~\bibnamefont {Bloch}},\ }\bibfield  {title} {\bibinfo {title} {A
  subradiant optical mirror formed by a single structured atomic layer},\
  }\href {https://doi.org/10.1038/s41586-020-2463-x} {\bibfield  {journal}
  {\bibinfo  {journal} {Nature}\ }\textbf {\bibinfo {volume} {583}},\ \bibinfo
  {pages} {369} (\bibinfo {year} {2020})}\BibitemShut {NoStop}%
\bibitem [{\citenamefont {Srakaew}\ \emph {et~al.}(2023)\citenamefont
  {Srakaew}, \citenamefont {Weckesser}, \citenamefont {Hollerith},
  \citenamefont {Wei}, \citenamefont {Adler}, \citenamefont {Bloch},\ and\
  \citenamefont {Zeiher}}]{Srakaew2023}%
  \BibitemOpen
  \bibfield  {author} {\bibinfo {author} {\bibfnamefont {K.}~\bibnamefont
  {Srakaew}}, \bibinfo {author} {\bibfnamefont {P.}~\bibnamefont {Weckesser}},
  \bibinfo {author} {\bibfnamefont {S.}~\bibnamefont {Hollerith}}, \bibinfo
  {author} {\bibfnamefont {D.}~\bibnamefont {Wei}}, \bibinfo {author}
  {\bibfnamefont {D.}~\bibnamefont {Adler}}, \bibinfo {author} {\bibfnamefont
  {I.}~\bibnamefont {Bloch}},\ and\ \bibinfo {author} {\bibfnamefont
  {J.}~\bibnamefont {Zeiher}},\ }\bibfield  {title} {\bibinfo {title} {A
  subwavelength atomic array switched by a single {Rydberg} atom},\ }\href
  {https://doi.org/10.1038/s41567-023-01959-y} {\bibfield  {journal} {\bibinfo
  {journal} {Nat. Phys.}\ }\textbf {\bibinfo {volume} {19}},\ \bibinfo {pages}
  {714} (\bibinfo {year} {2023})}\BibitemShut {NoStop}%
\bibitem [{\citenamefont {Douglas}\ \emph {et~al.}(2026)\citenamefont
  {Douglas}, \citenamefont {Su}, \citenamefont {Szurek}, \citenamefont {Groth},
  \citenamefont {Brandstetter}, \citenamefont {Markovi{\'c}}, \citenamefont
  {Rubies-Bigorda}, \citenamefont {Ostermann}, \citenamefont {Yelin},\ and\
  \citenamefont {Greiner}}]{Douglas2026}%
  \BibitemOpen
  \bibfield  {author} {\bibinfo {author} {\bibfnamefont {A.}~\bibnamefont
  {Douglas}}, \bibinfo {author} {\bibfnamefont {L.}~\bibnamefont {Su}},
  \bibinfo {author} {\bibfnamefont {M.}~\bibnamefont {Szurek}}, \bibinfo
  {author} {\bibfnamefont {R.}~\bibnamefont {Groth}}, \bibinfo {author}
  {\bibfnamefont {S.}~\bibnamefont {Brandstetter}}, \bibinfo {author}
  {\bibfnamefont {O.}~\bibnamefont {Markovi{\'c}}}, \bibinfo {author}
  {\bibfnamefont {O.}~\bibnamefont {Rubies-Bigorda}}, \bibinfo {author}
  {\bibfnamefont {S.}~\bibnamefont {Ostermann}}, \bibinfo {author}
  {\bibfnamefont {S.~F.}\ \bibnamefont {Yelin}},\ and\ \bibinfo {author}
  {\bibfnamefont {M.}~\bibnamefont {Greiner}},\ }\bibfield  {title} {\bibinfo
  {title} {Many-body super- and subradiance in ordered atomic arrays},\
  }\href@noop {} {\bibfield  {journal} {\bibinfo  {journal} {arXiv}\ }
  (\bibinfo {year} {2026})},\ \bibinfo {note} {arXiv:2604.11795}\BibitemShut
  {NoStop}%
\bibitem [{\citenamefont {Olmos}\ \emph {et~al.}(2013)\citenamefont {Olmos},
  \citenamefont {Yu}, \citenamefont {Singh}, \citenamefont {Schreck},
  \citenamefont {Bongs},\ and\ \citenamefont {Lesanovsky}}]{Olmos2013}%
  \BibitemOpen
  \bibfield  {author} {\bibinfo {author} {\bibfnamefont {B.}~\bibnamefont
  {Olmos}}, \bibinfo {author} {\bibfnamefont {D.}~\bibnamefont {Yu}}, \bibinfo
  {author} {\bibfnamefont {Y.}~\bibnamefont {Singh}}, \bibinfo {author}
  {\bibfnamefont {F.}~\bibnamefont {Schreck}}, \bibinfo {author} {\bibfnamefont
  {K.}~\bibnamefont {Bongs}},\ and\ \bibinfo {author} {\bibfnamefont
  {I.}~\bibnamefont {Lesanovsky}},\ }\bibfield  {title} {\bibinfo {title}
  {Long-range interacting many-body systems with alkaline-earth-metal atoms},\
  }\href {https://doi.org/10.1103/PhysRevLett.110.143602} {\bibfield  {journal}
  {\bibinfo  {journal} {Phys. Rev. Lett.}\ }\textbf {\bibinfo {volume} {110}},\
  \bibinfo {pages} {143602} (\bibinfo {year} {2013})}\BibitemShut {NoStop}%
\bibitem [{\citenamefont {Asenjo-Garcia}\ \emph
  {et~al.}(2017{\natexlab{a}})\citenamefont {Asenjo-Garcia}, \citenamefont
  {Moreno-Cardoner}, \citenamefont {Albrecht}, \citenamefont {Kimble},\ and\
  \citenamefont {Chang}}]{AsenjoGarcia2017}%
  \BibitemOpen
  \bibfield  {author} {\bibinfo {author} {\bibfnamefont {A.}~\bibnamefont
  {Asenjo-Garcia}}, \bibinfo {author} {\bibfnamefont {M.}~\bibnamefont
  {Moreno-Cardoner}}, \bibinfo {author} {\bibfnamefont {A.}~\bibnamefont
  {Albrecht}}, \bibinfo {author} {\bibfnamefont {H.~J.}\ \bibnamefont
  {Kimble}},\ and\ \bibinfo {author} {\bibfnamefont {D.~E.}\ \bibnamefont
  {Chang}},\ }\bibfield  {title} {\bibinfo {title} {Exponential improvement in
  photon storage fidelities using subradiance and ``{Selective Radiance}'' in
  atomic arrays},\ }\href {https://doi.org/10.1103/PhysRevX.7.031024}
  {\bibfield  {journal} {\bibinfo  {journal} {Phys. Rev. X}\ }\textbf {\bibinfo
  {volume} {7}},\ \bibinfo {pages} {031024} (\bibinfo {year}
  {2017}{\natexlab{a}})}\BibitemShut {NoStop}%
\bibitem [{\citenamefont {Reitz}\ \emph {et~al.}(2022)\citenamefont {Reitz},
  \citenamefont {Sommer},\ and\ \citenamefont {Genes}}]{Reitz2022}%
  \BibitemOpen
  \bibfield  {author} {\bibinfo {author} {\bibfnamefont {M.}~\bibnamefont
  {Reitz}}, \bibinfo {author} {\bibfnamefont {C.}~\bibnamefont {Sommer}},\ and\
  \bibinfo {author} {\bibfnamefont {C.}~\bibnamefont {Genes}},\ }\bibfield
  {title} {\bibinfo {title} {Cooperative quantum phenomena in light-matter
  platforms},\ }\href {https://doi.org/10.1103/PRXQuantum.3.010201} {\bibfield
  {journal} {\bibinfo  {journal} {PRX Quantum}\ }\textbf {\bibinfo {volume}
  {3}},\ \bibinfo {pages} {010201} (\bibinfo {year} {2022})}\BibitemShut
  {NoStop}%
\bibitem [{\citenamefont {Jen}\ \emph {et~al.}(2016)\citenamefont {Jen},
  \citenamefont {Chang},\ and\ \citenamefont {Chen}}]{Jen2016}%
  \BibitemOpen
  \bibfield  {author} {\bibinfo {author} {\bibfnamefont {H.~H.}\ \bibnamefont
  {Jen}}, \bibinfo {author} {\bibfnamefont {M.-S.}\ \bibnamefont {Chang}},\
  and\ \bibinfo {author} {\bibfnamefont {Y.-C.}\ \bibnamefont {Chen}},\
  }\bibfield  {title} {\bibinfo {title} {Cooperative single-photon subradiant
  states},\ }\href {https://doi.org/10.1103/PhysRevA.94.013803} {\bibfield
  {journal} {\bibinfo  {journal} {Phys. Rev. A}\ }\textbf {\bibinfo {volume}
  {94}},\ \bibinfo {pages} {013803} (\bibinfo {year} {2016})}\BibitemShut
  {NoStop}%
\bibitem [{\citenamefont {Jen}(2017)}]{Jen2017}%
  \BibitemOpen
  \bibfield  {author} {\bibinfo {author} {\bibfnamefont {H.~H.}\ \bibnamefont
  {Jen}},\ }\bibfield  {title} {\bibinfo {title} {Phase-imprinted multiphoton
  subradiant states},\ }\href {https://doi.org/10.1103/PhysRevA.96.023814}
  {\bibfield  {journal} {\bibinfo  {journal} {Phys. Rev. A}\ }\textbf {\bibinfo
  {volume} {96}},\ \bibinfo {pages} {023814} (\bibinfo {year}
  {2017})}\BibitemShut {NoStop}%
\bibitem [{\citenamefont {Needham}\ \emph {et~al.}(2019)\citenamefont
  {Needham}, \citenamefont {Lesanovsky},\ and\ \citenamefont
  {Olmos}}]{Needham2019}%
  \BibitemOpen
  \bibfield  {author} {\bibinfo {author} {\bibfnamefont {J.~A.}\ \bibnamefont
  {Needham}}, \bibinfo {author} {\bibfnamefont {I.}~\bibnamefont
  {Lesanovsky}},\ and\ \bibinfo {author} {\bibfnamefont {B.}~\bibnamefont
  {Olmos}},\ }\bibfield  {title} {\bibinfo {title} {Subradiance-protected
  excitation transport},\ }\href {https://doi.org/10.1088/1367-2630/ab2f44}
  {\bibfield  {journal} {\bibinfo  {journal} {New J. Phys.}\ }\textbf {\bibinfo
  {volume} {21}},\ \bibinfo {pages} {073061} (\bibinfo {year}
  {2019})}\BibitemShut {NoStop}%
\bibitem [{\citenamefont {Bettles}\ \emph {et~al.}(2016)\citenamefont
  {Bettles}, \citenamefont {Gardiner},\ and\ \citenamefont
  {Adams}}]{Bettles2016}%
  \BibitemOpen
  \bibfield  {author} {\bibinfo {author} {\bibfnamefont {R.~J.}\ \bibnamefont
  {Bettles}}, \bibinfo {author} {\bibfnamefont {S.~A.}\ \bibnamefont
  {Gardiner}},\ and\ \bibinfo {author} {\bibfnamefont {C.~S.}\ \bibnamefont
  {Adams}},\ }\bibfield  {title} {\bibinfo {title} {Enhanced optical cross
  section via collective coupling of atomic dipoles in a 2d array},\ }\href
  {https://doi.org/10.1103/PhysRevLett.116.103602} {\bibfield  {journal}
  {\bibinfo  {journal} {Phys. Rev. Lett.}\ }\textbf {\bibinfo {volume} {116}},\
  \bibinfo {pages} {103602} (\bibinfo {year} {2016})}\BibitemShut {NoStop}%
\bibitem [{\citenamefont {Bettles}\ \emph {et~al.}(2017)\citenamefont
  {Bettles}, \citenamefont {Min{\'a}{\v r}}, \citenamefont {Adams},
  \citenamefont {Lesanovsky},\ and\ \citenamefont {Olmos}}]{Bettles2017}%
  \BibitemOpen
  \bibfield  {author} {\bibinfo {author} {\bibfnamefont {R.~J.}\ \bibnamefont
  {Bettles}}, \bibinfo {author} {\bibfnamefont {J.}~\bibnamefont {Min{\'a}{\v
  r}}}, \bibinfo {author} {\bibfnamefont {C.~S.}\ \bibnamefont {Adams}},
  \bibinfo {author} {\bibfnamefont {I.}~\bibnamefont {Lesanovsky}},\ and\
  \bibinfo {author} {\bibfnamefont {B.}~\bibnamefont {Olmos}},\ }\bibfield
  {title} {\bibinfo {title} {Topological properties of a dense atomic lattice
  gas},\ }\href {https://doi.org/10.1103/PhysRevA.96.041603} {\bibfield
  {journal} {\bibinfo  {journal} {Phys. Rev. A}\ }\textbf {\bibinfo {volume}
  {96}},\ \bibinfo {pages} {041603(R)} (\bibinfo {year} {2017})}\BibitemShut
  {NoStop}%
\bibitem [{\citenamefont {Perczel}\ \emph {et~al.}(2017)\citenamefont
  {Perczel}, \citenamefont {Borregaard}, \citenamefont {Chang}, \citenamefont
  {Pichler}, \citenamefont {Yelin}, \citenamefont {Zoller},\ and\ \citenamefont
  {Lukin}}]{Perczel2017}%
  \BibitemOpen
  \bibfield  {author} {\bibinfo {author} {\bibfnamefont {J.}~\bibnamefont
  {Perczel}}, \bibinfo {author} {\bibfnamefont {J.}~\bibnamefont {Borregaard}},
  \bibinfo {author} {\bibfnamefont {D.~E.}\ \bibnamefont {Chang}}, \bibinfo
  {author} {\bibfnamefont {H.}~\bibnamefont {Pichler}}, \bibinfo {author}
  {\bibfnamefont {S.~F.}\ \bibnamefont {Yelin}}, \bibinfo {author}
  {\bibfnamefont {P.}~\bibnamefont {Zoller}},\ and\ \bibinfo {author}
  {\bibfnamefont {M.~D.}\ \bibnamefont {Lukin}},\ }\bibfield  {title} {\bibinfo
  {title} {Topological quantum optics in two-dimensional atomic arrays},\
  }\href {https://doi.org/10.1103/PhysRevLett.119.023603} {\bibfield  {journal}
  {\bibinfo  {journal} {Phys. Rev. Lett.}\ }\textbf {\bibinfo {volume} {119}},\
  \bibinfo {pages} {023603} (\bibinfo {year} {2017})}\BibitemShut {NoStop}%
\bibitem [{\citenamefont {Browaeys}\ and\ \citenamefont
  {Lahaye}(2020)}]{BrowaeysLahaye2020}%
  \BibitemOpen
  \bibfield  {author} {\bibinfo {author} {\bibfnamefont {A.}~\bibnamefont
  {Browaeys}}\ and\ \bibinfo {author} {\bibfnamefont {T.}~\bibnamefont
  {Lahaye}},\ }\bibfield  {title} {\bibinfo {title} {Many-body physics with
  individually controlled {Rydberg} atoms},\ }\href
  {https://doi.org/10.1038/s41567-019-0733-z} {\bibfield  {journal} {\bibinfo
  {journal} {Nat. Phys.}\ }\textbf {\bibinfo {volume} {16}},\ \bibinfo {pages}
  {132} (\bibinfo {year} {2020})}\BibitemShut {NoStop}%
\bibitem [{\citenamefont {Moreno-Cardoner}\ \emph {et~al.}(2019)\citenamefont
  {Moreno-Cardoner}, \citenamefont {Plankensteiner}, \citenamefont {Ostermann},
  \citenamefont {Chang},\ and\ \citenamefont {Ritsch}}]{MorenoCardoner2019}%
  \BibitemOpen
  \bibfield  {author} {\bibinfo {author} {\bibfnamefont {M.}~\bibnamefont
  {Moreno-Cardoner}}, \bibinfo {author} {\bibfnamefont {D.}~\bibnamefont
  {Plankensteiner}}, \bibinfo {author} {\bibfnamefont {L.}~\bibnamefont
  {Ostermann}}, \bibinfo {author} {\bibfnamefont {D.~E.}\ \bibnamefont
  {Chang}},\ and\ \bibinfo {author} {\bibfnamefont {H.}~\bibnamefont
  {Ritsch}},\ }\bibfield  {title} {\bibinfo {title} {Subradiance-enhanced
  excitation transfer between dipole-coupled nanorings of quantum emitters},\
  }\href {https://doi.org/10.1103/PhysRevA.100.023806} {\bibfield  {journal}
  {\bibinfo  {journal} {Phys. Rev. A}\ }\textbf {\bibinfo {volume} {100}},\
  \bibinfo {pages} {023806} (\bibinfo {year} {2019})}\BibitemShut {NoStop}%
\bibitem [{\citenamefont {Cremer}\ \emph {et~al.}(2020)\citenamefont {Cremer},
  \citenamefont {Plankensteiner}, \citenamefont {Moreno-Cardoner},
  \citenamefont {Ostermann},\ and\ \citenamefont {Ritsch}}]{Cremer2020}%
  \BibitemOpen
  \bibfield  {author} {\bibinfo {author} {\bibfnamefont {J.}~\bibnamefont
  {Cremer}}, \bibinfo {author} {\bibfnamefont {D.}~\bibnamefont
  {Plankensteiner}}, \bibinfo {author} {\bibfnamefont {M.}~\bibnamefont
  {Moreno-Cardoner}}, \bibinfo {author} {\bibfnamefont {L.}~\bibnamefont
  {Ostermann}},\ and\ \bibinfo {author} {\bibfnamefont {H.}~\bibnamefont
  {Ritsch}},\ }\bibfield  {title} {\bibinfo {title} {Polarization control of
  radiation and energy flow in dipole-coupled nanorings},\ }\href
  {https://doi.org/10.1088/1367-2630/aba4d4} {\bibfield  {journal} {\bibinfo
  {journal} {New J. Phys.}\ }\textbf {\bibinfo {volume} {22}},\ \bibinfo
  {pages} {083052} (\bibinfo {year} {2020})}\BibitemShut {NoStop}%
\bibitem [{\citenamefont {Jen}\ \emph {et~al.}(2018)\citenamefont {Jen},
  \citenamefont {Chang},\ and\ \citenamefont {Chen}}]{Jen2018a}%
  \BibitemOpen
  \bibfield  {author} {\bibinfo {author} {\bibfnamefont {H.~H.}\ \bibnamefont
  {Jen}}, \bibinfo {author} {\bibfnamefont {M.-S.}\ \bibnamefont {Chang}},\
  and\ \bibinfo {author} {\bibfnamefont {Y.-C.}\ \bibnamefont {Chen}},\
  }\bibfield  {title} {\bibinfo {title} {Cooperative light scattering from
  helical-phase-imprinted atomic rings},\ }\href
  {https://doi.org/10.1038/s41598-018-27888-y} {\bibfield  {journal} {\bibinfo
  {journal} {Sci. Rep.}\ }\textbf {\bibinfo {volume} {8}},\ \bibinfo {pages}
  {9570} (\bibinfo {year} {2018})}\BibitemShut {NoStop}%
\bibitem [{\citenamefont {Jen}(2018)}]{Jen2018b}%
  \BibitemOpen
  \bibfield  {author} {\bibinfo {author} {\bibfnamefont {H.~H.}\ \bibnamefont
  {Jen}},\ }\bibfield  {title} {\bibinfo {title} {Directional subradiance from
  helical-phase-imprinted multiphoton states},\ }\href
  {https://doi.org/10.1038/s41598-018-25592-5} {\bibfield  {journal} {\bibinfo
  {journal} {Sci. Rep.}\ }\textbf {\bibinfo {volume} {8}},\ \bibinfo {pages}
  {7163} (\bibinfo {year} {2018})}\BibitemShut {NoStop}%
\bibitem [{\citenamefont {Scheil}\ \emph {et~al.}(2023)\citenamefont {Scheil},
  \citenamefont {Holzinger}, \citenamefont {Moreno-Cardoner},\ and\
  \citenamefont {Ritsch}}]{Scheil2023}%
  \BibitemOpen
  \bibfield  {author} {\bibinfo {author} {\bibfnamefont {V.}~\bibnamefont
  {Scheil}}, \bibinfo {author} {\bibfnamefont {R.}~\bibnamefont {Holzinger}},
  \bibinfo {author} {\bibfnamefont {M.}~\bibnamefont {Moreno-Cardoner}},\ and\
  \bibinfo {author} {\bibfnamefont {H.}~\bibnamefont {Ritsch}},\ }\bibfield
  {title} {\bibinfo {title} {Optical properties of concentric nanorings of
  quantum emitters},\ }\href {https://doi.org/10.3390/nano13050851} {\bibfield
  {journal} {\bibinfo  {journal} {Nanomaterials}\ }\textbf {\bibinfo {volume}
  {13}},\ \bibinfo {pages} {851} (\bibinfo {year} {2023})}\BibitemShut
  {NoStop}%
\bibitem [{\citenamefont {Holzinger}\ \emph {et~al.}(2024)\citenamefont
  {Holzinger}, \citenamefont {Peter}, \citenamefont {Ostermann}, \citenamefont
  {Ritsch},\ and\ \citenamefont {Yelin}}]{Holzinger2024}%
  \BibitemOpen
  \bibfield  {author} {\bibinfo {author} {\bibfnamefont {R.}~\bibnamefont
  {Holzinger}}, \bibinfo {author} {\bibfnamefont {J.~S.}\ \bibnamefont
  {Peter}}, \bibinfo {author} {\bibfnamefont {S.}~\bibnamefont {Ostermann}},
  \bibinfo {author} {\bibfnamefont {H.}~\bibnamefont {Ritsch}},\ and\ \bibinfo
  {author} {\bibfnamefont {S.~F.}\ \bibnamefont {Yelin}},\ }\bibfield  {title}
  {\bibinfo {title} {Harnessing quantum emitter rings for efficient energy
  transport and trapping},\ }\href {https://doi.org/10.1364/OPTICAQ.510021}
  {\bibfield  {journal} {\bibinfo  {journal} {Optica Quantum}\ }\textbf
  {\bibinfo {volume} {2}},\ \bibinfo {pages} {57} (\bibinfo {year}
  {2024})}\BibitemShut {NoStop}%
\bibitem [{\citenamefont {Guimond}\ \emph {et~al.}(2019)\citenamefont
  {Guimond}, \citenamefont {Grankin}, \citenamefont {Vasilyev}, \citenamefont
  {Vermersch},\ and\ \citenamefont {Zoller}}]{Guimond2019}%
  \BibitemOpen
  \bibfield  {author} {\bibinfo {author} {\bibfnamefont {P.-O.}\ \bibnamefont
  {Guimond}}, \bibinfo {author} {\bibfnamefont {A.}~\bibnamefont {Grankin}},
  \bibinfo {author} {\bibfnamefont {D.~V.}\ \bibnamefont {Vasilyev}}, \bibinfo
  {author} {\bibfnamefont {B.}~\bibnamefont {Vermersch}},\ and\ \bibinfo
  {author} {\bibfnamefont {P.}~\bibnamefont {Zoller}},\ }\bibfield  {title}
  {\bibinfo {title} {Subradiant bell states in distant atomic arrays},\ }\href
  {https://doi.org/10.1103/PhysRevLett.122.093601} {\bibfield  {journal}
  {\bibinfo  {journal} {Phys. Rev. Lett.}\ }\textbf {\bibinfo {volume} {122}},\
  \bibinfo {pages} {093601} (\bibinfo {year} {2019})}\BibitemShut {NoStop}%
\bibitem [{\citenamefont {Manzoni}\ \emph {et~al.}(2018)\citenamefont
  {Manzoni}, \citenamefont {Moreno-Cardoner}, \citenamefont {Asenjo-Garcia},
  \citenamefont {Porto}, \citenamefont {Gorshkov},\ and\ \citenamefont
  {Chang}}]{Manzoni2018}%
  \BibitemOpen
  \bibfield  {author} {\bibinfo {author} {\bibfnamefont {M.~T.}\ \bibnamefont
  {Manzoni}}, \bibinfo {author} {\bibfnamefont {M.}~\bibnamefont
  {Moreno-Cardoner}}, \bibinfo {author} {\bibfnamefont {A.}~\bibnamefont
  {Asenjo-Garcia}}, \bibinfo {author} {\bibfnamefont {J.~V.}\ \bibnamefont
  {Porto}}, \bibinfo {author} {\bibfnamefont {A.~V.}\ \bibnamefont
  {Gorshkov}},\ and\ \bibinfo {author} {\bibfnamefont {D.~E.}\ \bibnamefont
  {Chang}},\ }\bibfield  {title} {\bibinfo {title} {Optimization of photon
  storage fidelity in ordered atomic arrays},\ }\href
  {https://doi.org/10.1088/1367-2630/aadb74} {\bibfield  {journal} {\bibinfo
  {journal} {New J. Phys.}\ }\textbf {\bibinfo {volume} {20}},\ \bibinfo
  {pages} {083048} (\bibinfo {year} {2018})}\BibitemShut {NoStop}%
\bibitem [{\citenamefont {Ballantine}\ and\ \citenamefont
  {Ruostekoski}(2021)}]{Ballantine2021}%
  \BibitemOpen
  \bibfield  {author} {\bibinfo {author} {\bibfnamefont {K.~E.}\ \bibnamefont
  {Ballantine}}\ and\ \bibinfo {author} {\bibfnamefont {J.}~\bibnamefont
  {Ruostekoski}},\ }\bibfield  {title} {\bibinfo {title} {Quantum single-photon
  control, storage, and entanglement generation with planar atomic arrays},\
  }\href {https://doi.org/10.1103/PRXQuantum.2.040362} {\bibfield  {journal}
  {\bibinfo  {journal} {PRX Quantum}\ }\textbf {\bibinfo {volume} {2}},\
  \bibinfo {pages} {040362} (\bibinfo {year} {2021})}\BibitemShut {NoStop}%
\bibitem [{\citenamefont {Ballantine}\ and\ \citenamefont
  {Ruostekoski}(2022)}]{Ballantine2022}%
  \BibitemOpen
  \bibfield  {author} {\bibinfo {author} {\bibfnamefont {K.~E.}\ \bibnamefont
  {Ballantine}}\ and\ \bibinfo {author} {\bibfnamefont {J.}~\bibnamefont
  {Ruostekoski}},\ }\bibfield  {title} {\bibinfo {title} {Unidirectional
  absorption, storage, and emission of single photons in a collectively
  responding bilayer atomic array},\ }\href
  {https://doi.org/10.1103/PhysRevResearch.4.033200} {\bibfield  {journal}
  {\bibinfo  {journal} {Phys. Rev. Res.}\ }\textbf {\bibinfo {volume} {4}},\
  \bibinfo {pages} {033200} (\bibinfo {year} {2022})}\BibitemShut {NoStop}%
\bibitem [{\citenamefont {Rubies-Bigorda}\ \emph {et~al.}(2022)\citenamefont
  {Rubies-Bigorda}, \citenamefont {Walther}, \citenamefont {Patti},\ and\
  \citenamefont {Yelin}}]{RubiesBigorda2022}%
  \BibitemOpen
  \bibfield  {author} {\bibinfo {author} {\bibfnamefont {O.}~\bibnamefont
  {Rubies-Bigorda}}, \bibinfo {author} {\bibfnamefont {V.}~\bibnamefont
  {Walther}}, \bibinfo {author} {\bibfnamefont {T.~L.}\ \bibnamefont {Patti}},\
  and\ \bibinfo {author} {\bibfnamefont {S.~F.}\ \bibnamefont {Yelin}},\
  }\bibfield  {title} {\bibinfo {title} {Photon control and coherent
  interactions via lattice dark states in atomic arrays},\ }\href
  {https://doi.org/10.1103/PhysRevResearch.4.013110} {\bibfield  {journal}
  {\bibinfo  {journal} {Phys. Rev. Res.}\ }\textbf {\bibinfo {volume} {4}},\
  \bibinfo {pages} {013110} (\bibinfo {year} {2022})}\BibitemShut {NoStop}%
\bibitem [{\citenamefont {Cech}\ \emph {et~al.}(2023)\citenamefont {Cech},
  \citenamefont {Lesanovsky},\ and\ \citenamefont {Olmos}}]{Cech2023}%
  \BibitemOpen
  \bibfield  {author} {\bibinfo {author} {\bibfnamefont {M.}~\bibnamefont
  {Cech}}, \bibinfo {author} {\bibfnamefont {I.}~\bibnamefont {Lesanovsky}},\
  and\ \bibinfo {author} {\bibfnamefont {B.}~\bibnamefont {Olmos}},\ }\bibfield
   {title} {\bibinfo {title} {Dispersionless subradiant photon storage in
  one-dimensional emitter chains},\ }\href
  {https://doi.org/10.1103/PhysRevA.108.L051702} {\bibfield  {journal}
  {\bibinfo  {journal} {Phys. Rev. A}\ }\textbf {\bibinfo {volume} {108}},\
  \bibinfo {pages} {L051702} (\bibinfo {year} {2023})}\BibitemShut {NoStop}%
\bibitem [{\citenamefont {Poshakinskiy}\ and\ \citenamefont
  {Poddubny}(2021)}]{Poshakinskiy2021}%
  \BibitemOpen
  \bibfield  {author} {\bibinfo {author} {\bibfnamefont {A.~V.}\ \bibnamefont
  {Poshakinskiy}}\ and\ \bibinfo {author} {\bibfnamefont {A.~N.}\ \bibnamefont
  {Poddubny}},\ }\bibfield  {title} {\bibinfo {title} {Dimerization of
  many-body subradiant states in waveguide quantum electrodynamics},\ }\href
  {https://doi.org/10.1103/PhysRevLett.127.173601} {\bibfield  {journal}
  {\bibinfo  {journal} {Phys. Rev. Lett.}\ }\textbf {\bibinfo {volume} {127}},\
  \bibinfo {pages} {173601} (\bibinfo {year} {2021})}\BibitemShut {NoStop}%
\bibitem [{\citenamefont {Asenjo-Garcia}\ \emph
  {et~al.}(2017{\natexlab{b}})\citenamefont {Asenjo-Garcia}, \citenamefont
  {Hood}, \citenamefont {Chang},\ and\ \citenamefont
  {Kimble}}]{AsenjoGarcia2017a}%
  \BibitemOpen
  \bibfield  {author} {\bibinfo {author} {\bibfnamefont {A.}~\bibnamefont
  {Asenjo-Garcia}}, \bibinfo {author} {\bibfnamefont {J.~D.}\ \bibnamefont
  {Hood}}, \bibinfo {author} {\bibfnamefont {D.~E.}\ \bibnamefont {Chang}},\
  and\ \bibinfo {author} {\bibfnamefont {H.~J.}\ \bibnamefont {Kimble}},\
  }\bibfield  {title} {\bibinfo {title} {Atom-light interactions in
  quasi-one-dimensional nanostructures: A green's-function perspective},\
  }\href {https://doi.org/10.1103/PhysRevA.95.033818} {\bibfield  {journal}
  {\bibinfo  {journal} {Phys. Rev. A}\ }\textbf {\bibinfo {volume} {95}},\
  \bibinfo {pages} {033818} (\bibinfo {year} {2017}{\natexlab{b}})}\BibitemShut
  {NoStop}%
\bibitem [{\citenamefont {Jones}\ \emph {et~al.}(2018)\citenamefont {Jones},
  \citenamefont {Needham}, \citenamefont {Lesanovsky}, \citenamefont
  {Intravaia},\ and\ \citenamefont {Olmos}}]{Jones2018}%
  \BibitemOpen
  \bibfield  {author} {\bibinfo {author} {\bibfnamefont {R.}~\bibnamefont
  {Jones}}, \bibinfo {author} {\bibfnamefont {J.~A.}\ \bibnamefont {Needham}},
  \bibinfo {author} {\bibfnamefont {I.}~\bibnamefont {Lesanovsky}}, \bibinfo
  {author} {\bibfnamefont {F.}~\bibnamefont {Intravaia}},\ and\ \bibinfo
  {author} {\bibfnamefont {B.}~\bibnamefont {Olmos}},\ }\bibfield  {title}
  {\bibinfo {title} {Modified dipole-dipole interaction and dissipation in an
  atomic ensemble near surfaces},\ }\href
  {https://doi.org/10.1103/PhysRevA.97.053841} {\bibfield  {journal} {\bibinfo
  {journal} {Phys. Rev. A}\ }\textbf {\bibinfo {volume} {97}},\ \bibinfo
  {pages} {053841} (\bibinfo {year} {2018})}\BibitemShut {NoStop}%
\bibitem [{Sup()}]{SupMat}%
  \BibitemOpen
  \href@noop {} {}\bibinfo {note} {See Supplemental Material for a more
  detailed analysis.}\BibitemShut {Stop}%
\bibitem [{\citenamefont {Leppenen}\ and\ \citenamefont
  {Poddubny}(2026)}]{Leppenen2026}%
  \BibitemOpen
  \bibfield  {author} {\bibinfo {author} {\bibfnamefont {N.}~\bibnamefont
  {Leppenen}}\ and\ \bibinfo {author} {\bibfnamefont {A.~N.}\ \bibnamefont
  {Poddubny}},\ }\bibfield  {title} {\bibinfo {title} {Persistent subradiant
  correlations in a disordered ensemble of collectively driven quantum
  emitters},\ }\href {https://doi.org/10.1103/czf3-74sf} {\bibfield  {journal}
  {\bibinfo  {journal} {Phys. Rev. A}\ }\textbf {\bibinfo {volume} {113}},\
  \bibinfo {pages} {023709} (\bibinfo {year} {2026})}\BibitemShut {NoStop}%
\bibitem [{\citenamefont {Tiarks}\ \emph {et~al.}(2016)\citenamefont {Tiarks},
  \citenamefont {Schmidt}, \citenamefont {Rempe},\ and\ \citenamefont
  {D{\"u}rr}}]{Tiarks2016}%
  \BibitemOpen
  \bibfield  {author} {\bibinfo {author} {\bibfnamefont {D.}~\bibnamefont
  {Tiarks}}, \bibinfo {author} {\bibfnamefont {S.}~\bibnamefont {Schmidt}},
  \bibinfo {author} {\bibfnamefont {G.}~\bibnamefont {Rempe}},\ and\ \bibinfo
  {author} {\bibfnamefont {S.}~\bibnamefont {D{\"u}rr}},\ }\bibfield  {title}
  {\bibinfo {title} {Optical {$\pi$} phase shift created with a single-photon
  pulse},\ }\href {https://doi.org/10.1126/sciadv.1600036} {\bibfield
  {journal} {\bibinfo  {journal} {Science Advances}\ }\textbf {\bibinfo
  {volume} {2}},\ \bibinfo {pages} {e1600036} (\bibinfo {year}
  {2016})}\BibitemShut {NoStop}%
\bibitem [{\citenamefont {Du}\ \emph {et~al.}(2026)\citenamefont {Du},
  \citenamefont {Guo},\ and\ \citenamefont {Ma}}]{Du2026}%
  \BibitemOpen
  \bibfield  {author} {\bibinfo {author} {\bibfnamefont {B.}~\bibnamefont
  {Du}}, \bibinfo {author} {\bibfnamefont {Q.}~\bibnamefont {Guo}},\ and\
  \bibinfo {author} {\bibfnamefont {R.}~\bibnamefont {Ma}},\ }\href
  {https://arxiv.org/abs/2605.12442} {\bibinfo {title} {Programmable
  superradiance in an interacting qubit array}} (\bibinfo {year} {2026}),\
  \Eprint {https://arxiv.org/abs/2605.12442} {arXiv:2605.12442
  [cond-mat.quant-gas]} \BibitemShut {NoStop}%
\bibitem [{\citenamefont {Holman}\ \emph {et~al.}(2026)\citenamefont {Holman},
  \citenamefont {Sun}, \citenamefont {Seo}, \citenamefont {Corn}, \citenamefont
  {Zhu}, \citenamefont {Xu}, \citenamefont {Wu}, \citenamefont {Yu},
  \citenamefont {Filin}, \citenamefont {Safronova},\ and\ \citenamefont
  {Will}}]{Holman2026}%
  \BibitemOpen
  \bibfield  {author} {\bibinfo {author} {\bibfnamefont {A.}~\bibnamefont
  {Holman}}, \bibinfo {author} {\bibfnamefont {X.}~\bibnamefont {Sun}},
  \bibinfo {author} {\bibfnamefont {B.}~\bibnamefont {Seo}}, \bibinfo {author}
  {\bibfnamefont {J.}~\bibnamefont {Corn}}, \bibinfo {author} {\bibfnamefont
  {Z.}~\bibnamefont {Zhu}}, \bibinfo {author} {\bibfnamefont {Y.}~\bibnamefont
  {Xu}}, \bibinfo {author} {\bibfnamefont {J.}~\bibnamefont {Wu}}, \bibinfo
  {author} {\bibfnamefont {N.}~\bibnamefont {Yu}}, \bibinfo {author}
  {\bibfnamefont {D.}~\bibnamefont {Filin}}, \bibinfo {author} {\bibfnamefont
  {M.}~\bibnamefont {Safronova}},\ and\ \bibinfo {author} {\bibfnamefont
  {S.}~\bibnamefont {Will}},\ }\href {https://arxiv.org/abs/2606.02560}
  {\bibinfo {title} {A mid-infrared platform based on strontium tweezer
  arrays}} (\bibinfo {year} {2026}),\ \Eprint
  {https://arxiv.org/abs/2606.02560} {arXiv:2606.02560 [physics.atom-ph]}
  \BibitemShut {NoStop}%
\bibitem [{\citenamefont {Olmos}\ and\ \citenamefont
  {Lesanovsky}(2025)}]{Olmos2025}%
  \BibitemOpen
  \bibfield  {author} {\bibinfo {author} {\bibfnamefont {B.}~\bibnamefont
  {Olmos}}\ and\ \bibinfo {author} {\bibfnamefont {I.}~\bibnamefont
  {Lesanovsky}},\ }\bibfield  {title} {\bibinfo {title} {Hybrid sub- and
  superradiant states in emitter arrays with quantized motion},\ }\href
  {https://doi.org/10.1103/q2kj-w3lf} {\bibfield  {journal} {\bibinfo
  {journal} {Phys. Rev. Lett.}\ }\textbf {\bibinfo {volume} {134}},\ \bibinfo
  {pages} {243602} (\bibinfo {year} {2025})}\BibitemShut {NoStop}%
\bibitem [{\citenamefont {Rubies-Bigorda}\ \emph {et~al.}(2025)\citenamefont
  {Rubies-Bigorda}, \citenamefont {Holzinger}, \citenamefont {Asenjo-Garcia},
  \citenamefont {Romero-Isart}, \citenamefont {Ritsch}, \citenamefont
  {Ostermann}, \citenamefont {Gonzalez-Ballestero}, \citenamefont {Yelin},\
  and\ \citenamefont {Rusconi}}]{Rubies2025}%
  \BibitemOpen
  \bibfield  {author} {\bibinfo {author} {\bibfnamefont {O.}~\bibnamefont
  {Rubies-Bigorda}}, \bibinfo {author} {\bibfnamefont {R.}~\bibnamefont
  {Holzinger}}, \bibinfo {author} {\bibfnamefont {A.}~\bibnamefont
  {Asenjo-Garcia}}, \bibinfo {author} {\bibfnamefont {O.}~\bibnamefont
  {Romero-Isart}}, \bibinfo {author} {\bibfnamefont {H.}~\bibnamefont
  {Ritsch}}, \bibinfo {author} {\bibfnamefont {S.}~\bibnamefont {Ostermann}},
  \bibinfo {author} {\bibfnamefont {C.}~\bibnamefont {Gonzalez-Ballestero}},
  \bibinfo {author} {\bibfnamefont {S.~F.}\ \bibnamefont {Yelin}},\ and\
  \bibinfo {author} {\bibfnamefont {C.~C.}\ \bibnamefont {Rusconi}},\
  }\bibfield  {title} {\bibinfo {title} {Collectively enhanced ground-state
  cooling in subwavelength atomic arrays},\ }\href
  {https://doi.org/10.1103/bhwv-ndtj} {\bibfield  {journal} {\bibinfo
  {journal} {Phys. Rev. A}\ }\textbf {\bibinfo {volume} {112}},\ \bibinfo
  {pages} {023714} (\bibinfo {year} {2025})}\BibitemShut {NoStop}%
\end{thebibliography}%

\clearpage
\onecolumngrid
\setcounter{figure}{0}
\setcounter{equation}{0}
\setcounter{table}{0}
\renewcommand{\thefigure}{S\arabic{figure}}
\renewcommand{\theequation}{S\arabic{equation}}
\renewcommand{\thetable}{S\arabic{table}}
\renewcommand{\theHfigure}{S\arabic{figure}}
\renewcommand{\theHequation}{S\arabic{equation}}
\renewcommand{\theHtable}{S\arabic{table}}

\onecolumngrid

\begin{center}
  {\large\bfseries Supplemental Material:\\[0.3em]
   Coherent control of subradiant excitations in atomic rings\par}
  \vspace{0.8em}
  {Milena Djatchkova,$^{1}$ Igor Lesanovsky,$^{1, 2}$ and Beatriz Olmos$^{1}$\par}
  \vspace{0.4em}
  {\small\itshape
   $^{1}$Institut für Theoretische Physik, Universität Tübingen,\\
   Auf der Morgenstelle 14,
   72076 Tübingen, Germany\\
   $^{2}$School of Physics and Astronomy and Centre for the Mathematics \\
   and Theoretical Physics of Quantum Non-Equilibrium Systems, \\
   The University of Nottingham, Nottingham, NG7 2RD, United Kingdom}
\end{center}

\vspace{0.8em}

\begin{center}
\begin{minipage}{0.86\textwidth}
\noindent This Supplemental Material provides additional details supporting the results of the main text. We first analyze the hybridization of the localized eigenmodes of two rings and the effective beam-splitter-like behavior of the symmetric configuration. Furthermore, we outline the derivation of the equation of motion in the two-excitation case and the definition of the excess phase, demonstrate that the reported phenomena persist at larger interatomic spacings, quantify the robustness of the inter-ring transfer against misalignment of the rings, and document the numerical convergence of the reported values.
\end{minipage}
\end{center}

\vspace{1.2em}

\twocolumngrid

\begin{figure*}
    \centering
    \includegraphics[width=\textwidth]{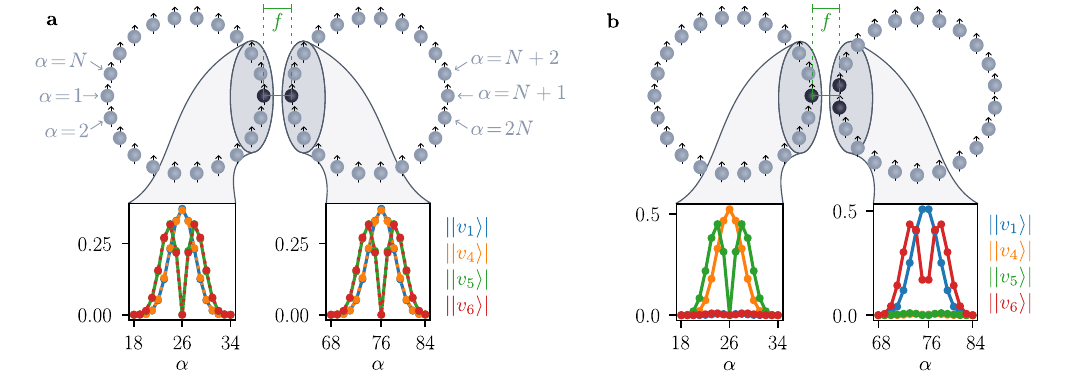}
    \caption{\textbf{a} Schematic of the symmetric configuration, with insets showing the spatial profiles of the lowest eigenstates of $H^\mathrm{eff}$ localized within the transfer region for $N=50$ atoms per ring. These eigenstates hybridize across both rings, where $\alpha$ denotes the continuous lattice site labeling, enabling excitation transfer between them. \textbf{b} Schematic of the shifted configuration, where the corresponding eigenstates instead localize on the individual rings, effectively decoupling them and suppressing excitation transfer. }
    \label{fig:fig2_3}
\end{figure*}

\begin{figure}
    \centering
    \includegraphics[width=0.5\textwidth]{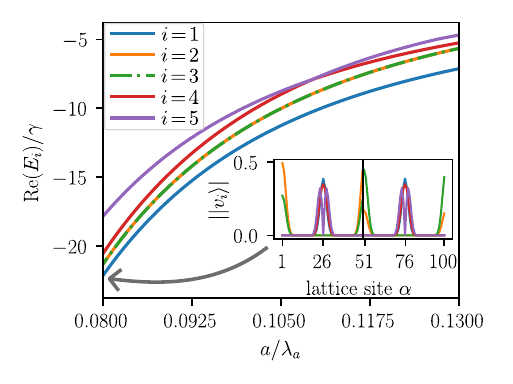}
    \caption{Real part of the eigenvalues $\mathrm{Re}(E_i)$ of the effective Hamiltonian for the symmetric configuration, representing the energy in units of the single-particle decay rate $\gamma$, as a function of the interatomic spacing $a$ in units of the wavelength $\lambda_a$. The indices $i = 1, \dots, 5$ label the five lowest-lying eigenmodes, and the ring separation is fixed at $f = 0.12\:\lambda_a$. The inset shows the spatial profile of these eigenmodes across the lattice sites $\alpha$, where $\alpha \leq 50$ corresponds to the first ring and $\alpha > 50$ to the second ring, evaluated at an interatomic spacing of $a = 0.08\:\lambda_a$.}
    \label{fig:encross}
\end{figure}

\begin{figure*}
    \centering
    \includegraphics[width=\textwidth]{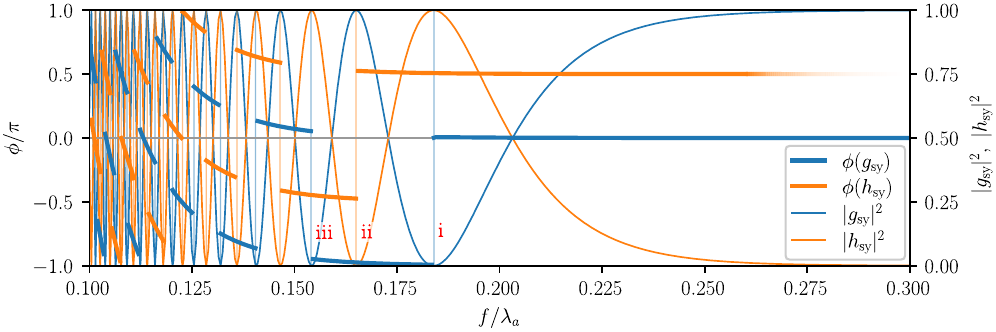}
    \caption{Squared magnitudes ($|g_\mathrm{sy}|^2$, $|h_\mathrm{sy}|^2$) and complex phases ($\phi(g_\mathrm{sy})$, $\phi(h_\mathrm{sy})$) of the single-excitation population fractions after a full $2\pi$ dipole rotation, plotted against ring separation $f$ for a fixed atomic spacing $a=0.08\:\lambda_a$ in the symmetric configuration. Vertical lines indicate the specific ring separations where the excitation is fully localized on a single ring, i.e., $g_\mathrm{sy}=0$ (blue) or $h_\mathrm{sy}=0$ (orange). After each full population transfer the respective phase exhibits a $\pi$-jump. Markers \textbf{i-iii} highlight the ring separations ($0.185\:\lambda_a, 0.165\:\lambda_a, 0.155\:\lambda_a$) for which exemplary time evolutions with one, two and three complete transfers, respectively, were shown in Fig.~\ref{fig2}.}
    \label{fig:gh}
\end{figure*}

\textit{Hybridization of localized eigenmodes ---} For the symmetric configuration of two sufficiently close rings, a single excitation can transfer from one ring to the other. Moreover, such transfers can happen multiple times. When the dipoles are aligned perpendicular to the inter-ring axis, see Fig.\:\ref{fig:fig2_3}a, all but two of the six lowest eigenmodes of the effective Hamiltonian $H^\mathrm{eff}$ are localized within the transfer area and hybridize across both rings, as can be seen in the insets. The remaining modes, that are localized away from the transfer region, play no role in the transfer dynamics and are not considered here. When the Gaussian excitation is brought into the transfer area via the adiabatic dipole rotation, it overlaps strongly with these ring-shared eigenstates, leading to a shape- and subradiance-preserving transfer onto the opposite ring. Since these eigenmodes are symmetric, strong inter-ring coupling or extended times in the transfer area can give rise to multiple back-and-forth transfers. Partial transfers are also possible and the resulting distribution of the excitation across both rings after the wave packet leaves the transfer area is investigated in detail below.

The discontinuity highlighted in orange in Fig.~\ref{fig2}c in the main text can be understood by further analyzing the low-lying eigenvalues of the effective Hamiltonian. In Fig.~\ref{fig:encross}, the real part of the eigenvalues $\mathrm{Re}(E_i)$, corresponding to the energy of the modes, is plotted as a function of the interatomic distance $a$ for a representative fixed ring separation $f=0.12\:\lambda_a$. An energy crossing of the fourth and fifth eigenvalues can be observed at the point of discontinuity. The initial wave function is constructed to minimize energy, since the low-lying modes are the most subradiant ones. Therefore the initial state acquires a large overlap with the fifth eigenmode after the energy crossing, rather than the fourth. The fourth eigenmode $\ket{v_4}$ exhibits a single peak near the central atom of the transfer area (see inset of Fig.~\ref{fig:encross}), whereas the fifth eigenmode $\ket{v_5}$ is double-peaked with a local minimum at its center. The larger overlap with this double-peaked eigenmode leads to a broadened wave function after the crossing. A broader wave function, in turn, requires smaller ring separations, so that the atoms further from the center come close enough to enable transfer, explaining the shift of the fringe pattern toward smaller ring separations in Fig.~\ref{fig2}c. In the vicinity of the energy crossing, the eigenvalues are nearly degenerate, thus the wave function couples simultaneously to both the fourth and fifth eigenmodes. Since the two eigenmodes require different ring separations for equivalent transfer dynamics, only partial transfer is possible in this regime, giving rise to the blurred region around the energy crossing.

The shifted configuration, with the dipoles again aligned perpendicular to the inter-ring axis, is depicted in Fig.~\ref{fig:fig2_3}b. The insets show the spatial profiles of the lowest relevant eigenmodes, each of which is localized on a single ring. As the overlap of the trapped excitation with eigenmodes localized on the opposite ring is negligible, the two rings are effectively decoupled, suppressing excitation transfer between them.

\textit{Effective beam splitter operator.---}
We now analyze multiple and partial excitation transfers in the symmetric configuration in more detail. To quantify them, we consider the overlap of the final wave function with the two initial states: $\ket{10}$ denotes the single excitation centered at the bottom of the left ring, and $\ket{01}$ the excitation centered at the top of the right ring, see Fig.~\ref{fig3}c in the main text. We define
\begin{align}
    g_\mathrm{sy} = \bra{10} \psi(t_\mathrm{end})\rangle, \quad h_\mathrm{sy} = \bra{01} \psi(t_\mathrm{end}) \rangle .
\end{align}
The magnitudes $|g_\mathrm{sy}|^2$ and $|h_\mathrm{sy}|^2$ give the fractions of the excitation located at the initial positions on the left and right ring, respectively. In general, these overlaps only provide a lower bound on the survival probability, $|g_\mathrm{sy}|^2 + |h_\mathrm{sy}|^2 \leq P_\mathrm{sur}$, since any distortion of the wave packets reduces their overlap with the initial states. In our case, however, both sides of this inequality nearly coincide, $|g_\mathrm{sy}|^2 + |h_\mathrm{sy}|^2 \approx P_\mathrm{sur}(t_\mathrm{end})$, with a maximal deviation below $10^{-3}$ over the entire range of ring separations considered. This nontrivial result reflects the remarkably faithful shape preservation during the adiabatic dipole rotation: almost the entire surviving excitation retains the shape of the two initial states and is merely redistributed between the rings. As shown in Fig.~\ref{fig:gh}, both quantities oscillate as the ring separation $f$ is decreased. They can be interpreted as analogues of the transmission and reflection coefficients of a beam splitter.

To extend this analogy to the acquired phases, we define
\begin{align}
    \phi(g_\mathrm{sy}) = \arg(g_\mathrm{sy}) - \phi_{1r}, \quad \phi(h_\mathrm{sy}) = \arg(h_\mathrm{sy}) - \phi_{1r},
\end{align}
where $\phi_{1r}$ is the free-evolution phase that a single excitation in an isolated ring acquires during the rotation protocol. Subtracting $\phi_{1r}$ isolates the phase imprinted solely by the two rings influencing each other. Most notably, whenever (part of) the excitation is transferred from one ring to the other, the transferred fraction acquires a phase shift of $\frac{\pi}{2}$ relative to the fraction that remains. In addition, the entire wave function acquires a global phase $\phi(f)$ caused by the mutual coupling of atoms in the two rings when these are in close proximity. A single partial transfer can thus be described as $\ket{10} \rightarrow e^{i\phi(f)}\left(|g|\ket{10} + i|h|\ket{01}\right)$. The symmetry of the configuration renders $\ket{01}$ and $\ket{10}$ interchangeable, so this operation proceeds identically in the reverse direction. For multiple transfers, each fraction accumulates a phase of $\frac{\pi}{2}$ per transfer.

Consider first a single complete transfer from the left to the right ring, which occurs at a ring separation of $f_\mathrm{i} = 0.185\,\lambda_a$, such that $|h(f_\mathrm{i})| = 1$ and $|g(f_\mathrm{i})| = 0$. For ring separations smaller than $f_\mathrm{i}$, i.e., for stronger inter-ring coupling, part of the wave function is transferred back to the initial ring and thus undergoes two transfers. This fraction therefore acquires the transfer phase twice, $\phi(g) = 2\cdot\frac{\pi}{2} \equiv -\pi \pmod{2\pi}$, which explains the phase jump around $f_\mathrm{i}$ (marked as i in Fig.~\ref{fig:gh}). The same effect occurs for $\phi(h)$ at $f_\mathrm{ii} = 0.165\,\lambda_a$. For ring separations slightly smaller than $f_\mathrm{ii}$, a fraction of the wave packet is transferred three times, resulting in $\phi(h) = 3\cdot\frac{\pi}{2} \equiv -\frac{\pi}{2} \pmod{2\pi}$. This pattern continues towards smaller ring separations, with a phase jump each time $|g| = 0$ or $|h| = 0$, i.e., whenever complete transfers occur. Denoting the number of complete transfers by $j$, the single-excitation state in the symmetric configuration after the rotation protocol can be described as
\begin{align}
    \ket{10} \rightarrow e^{i\phi(f)} \left( i^{2\lceil j/2 \rceil}\, |g|\ket{10} + i^{2\lfloor j/2 \rfloor +1}\, |h|\ket{01} \right).
\end{align}
The exponents correspond to the number of transfers undergone by the respective fraction, which is always even for $\ket{10}$ and odd for $\ket{01}$.

\begin{figure}[t]
    \centering
    \includegraphics[width=0.5\textwidth]{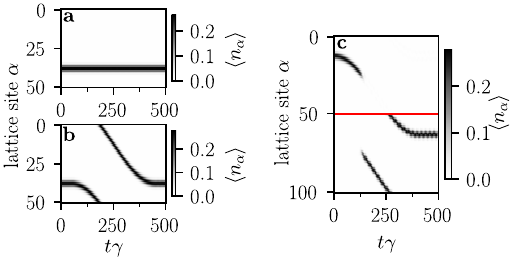}
    \vspace{-20pt}
    \caption{Representative excitation dynamics for rings with interatomic distance $a = 0.2\,\lambda_a$. \textbf{a} With constant dipole moments, a single excitation in one ring stays strongly localized. \textbf{b} Adiabatically rotating the dipoles over the time $\tau=400\:\gamma^{-1}$ transports the same excitation around the ring. \textbf{c} For two rings in the symmetric configuration at separation $f=0.29\:\lambda_a$, the excitation is transferred between rings. The survival probability remains close to one in all cases and the transfer efficiency in \textbf{c} is $P_\mathrm{tr} = 0.98$.
    }
    \label{fig:a0.2}
\end{figure}

\textit{Equation of Motion for Two Excitations.---}
For a single excitation, restricting to the excitation-preserving subspace and assuming an initially pure state, Eq.~\eqref{Lindblad-master} of the main text reduces to a Schrödinger-like differential equation~\cite{Needham2019,Cech2023}. We now derive the analogous result for two excitations. Under the same assumptions the state takes the form $\ket{\psi} = \sum_{i \neq j} c_{ij}\ket{e_i e_j}$, where the restriction $i \neq j$ excludes doubly excited atoms and the coefficients can be chosen symmetric, $c_{ij} = c_{ji}$. We collect them in a symmetric matrix $c(t)$ with entries $[c(t)]_{ij} = c_{ij}(t)$. Since the system is not driven, an excitation lost to the environment cannot be restored. The jump terms $\sigma_\alpha\rho\,\sigma_\beta^\dagger$ lower the excitation number on both sides of $\rho$, so that, with $\rho_{nm}=P_n\rho P_m$ the blocks of the density matrix between the sectors with $n$ and $m$ excitations, they feed $\rho_{22}$ only from $\rho_{33}$, which remains empty. The two-excitation block therefore obeys exactly
\begin{equation}\label{eq:rho22}
  \dot\rho_{22}=-\frac{i}{\hbar}\left(H^\mathrm{eff}\rho_{22}
               -\rho_{22}H^{\mathrm{eff}\dagger}\right),
\end{equation}
and, for a pure initial state, $\rho_{22}(t)=|\psi(t)\rangle\langle\psi(t)|$, with $|\psi(t)\rangle$ evolving under $H^\mathrm{eff}$. The non-Hermitian description is thus exact for all quantities evaluated within this sector. The norm, $P_\mathrm{sur}=\langle\psi|\psi\rangle$, is the probability that no photon has been emitted, and hence its decay constitutes a loss rather than an approximation error. It remains low throughout our analysis, except at the smallest ring separations in the symmetric configuration (see Fig. \ref{fig3}a of the main text). Correspondingly, as the Liouvillian is block triangular in the excitation numbers, its spectrum in this sector consists of the eigenvalues $-i(E_k-E_l^{*})/\hbar$, with $E_k$ the eigenvalues of $H^\mathrm{eff}$ in $\mathcal{H}_2$, so that its long-lived eigenmodes \cite{Leppenen2026} are formed by subradiant two-excitation eigenstates of $H^\mathrm{eff}$. Inserting the chosen ansatz into Eq.~\eqref{eq:rho22} and using the symmetry of $H^\mathrm{eff}$ yields
\begin{align}\label{eq:c2t}
    \frac{d}{dt} c(t) = -\frac{i}{\hbar} \Big( H^\mathrm{eff} c(t)
    + c(t)\, H^\mathrm{eff}\Big), \quad c_{ii}(t) = 0 \;\; \forall\, i, t.
\end{align}
The diagonal entries, which would correspond to the doubly excited states excluded above, are formally retained but constrained to zero. This allows us to write the equation of motion in a compact form. The dynamics of the two-excitation amplitudes is thus governed entirely by matrix products with the effective Hamiltonian.

\textit{Dependence on interatomic spacing.---} Subradiant trapping, intra-ring transport, and inter-ring transfer all persist at larger interatomic distances and ring separations, at the cost of reduced efficiency. Fig.\:\ref{fig:a0.2} shows these three processes for $a=0.2\:\lambda_a$. After the transfer between two rings at a distance $f=0.29\: \lambda_a$, the excitation exhibits visible oscillations within the trapping potential of the second ring, indicating that the wave packet acquires overlap with energetically higher-lying, less subradiant modes. This is a consequence of the imperfect coupling between the rings at this larger interatomic separation. Nevertheless, the transfer still reaches $P_\mathrm{tr} = 0.9823$ and the survival probability remains high throughout, with $P_\mathrm{sur}(t_\mathrm{end})>0.9999$ for the single ring and $0.9996$ for the two-ring configuration. The phenomena thus remain robust against increased distances, provided the lattice stays subwavelength.

\begin{figure}[t]
    \centering
    \includegraphics[width=0.48\textwidth]{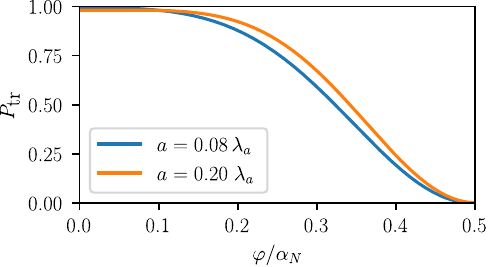}
    \vspace{-20pt}
    \caption{ Transfer probability $P_\text{tr}$ after one full dipole rotation as a function of the rotation angle $\varphi$ of the right ring, given in units of the angular spacing $\alpha_N = \frac{2 \pi}{N}$ between neighboring atoms, for the interatomic spacing $a=0.08 \, \lambda_a$ and ring separation $f = 0.185 \, \lambda_a$ (blue) and $a = 0.2 \, \lambda_a$, $f = 0.29\,\lambda_a$ (orange). At $\varphi = 0$ the two rings are aligned (symmetric configuration) and for $\varphi = \frac{\alpha_N}{2}$ the shifted configuration is recovered. }
    \label{fig:robustness}
\end{figure}

\textit{Robustness against ring misalignment.} --- For an experimental realization, the sensitivity of the transfer to the relative alignment of the two rings is an important figure of merit. We consider the symmetric configuration with parameters chosen such that exactly one full coherent transfer between the rings takes place. Each ring consists of $N=50$ atoms, so that neighboring atoms are separated by the angle $\alpha_N=2\pi/N$. Rotating the right ring by $\varphi=\alpha_N/2$ yields the shifted configuration, in which the transfer is completely suppressed. In Fig.~\ref{fig:robustness} we vary the rotation angle of the right ring in the range $0\leq\varphi\leq\alpha_N/2$. For both interatomic spacings, $a=0.08\,\lambda_\mathrm{a}$ and $a=0.2\,\lambda_\mathrm{a}$, the transfer probability exhibits a broad plateau around $\varphi=0$ before it decreases gradually and vanishes at $\varphi=\alpha_N/2$. It remains above $P_\mathrm{tr}=0.95$ for $\varphi\leq0.145\,\alpha_N$, which corresponds to a displacement of the atoms at the point of closest approach by $0.145\,a$. Since a lateral displacement of one ring along the tangent at this point is locally equivalent to a rotation, the same tolerance applies to it. The dependence on the ring separation is shown in Fig.~\ref{fig2}c of the main text. Near a full-transfer point, $P_\mathrm{tr}$ is maximal as a function of $f$ and hence insensitive to small deviations to first order. %

\textit{Two-excitation dynamics and the excess phase.} --- The reference state $\ket{\psi_\mathrm{ind}(t)}$ entering Eq.~\eqref{eq:phase} of the main text is obtained by integrating Eq.~\eqref{eq:c2t} without the constraint $c_{ii}=0$. Since the initial state contains no doubly excited components, this unconstrained evolution is exactly the symmetrized product of two single-excitation evolutions. This reference is required because the proximity of the two rings itself affects the acquired phase, as already observed in the single-excitation case. Without this influence, the phases in Fig.~\ref{fig:gh} would remain constant apart from the $\frac{\pi}{2}$ jumps. Instead, they increase with decreasing ring separation. To isolate the interaction-induced phase and remove the contribution arising from the mere proximity of the rings, we take the difference between the phases obtained in the interacting and non-interacting cases.

In the shifted configuration, this difference directly yields the desired interaction phase. In the symmetric configuration, by contrast, the phases of $n\pi/2$ acquired through the inter-ring transfers must be taken into account. Because two excitations are involved, the total phase changes by integer multiples of $\pi$. Analogous to the Hong-Ou-Mandel effect, these $\pi$ jumps occur at ring separations where the excitation is equally distributed between the two rings, $|g(f)| = |h(f)|$ (cf.~Fig.~\ref{fig:gh}). These jumps arise from the sign change of the reference amplitude and are absorbed into the exchange contribution $\Delta\phi_\text{sy}^\text{ex}$, equal to $\pi$ where the excitations are predominantly exchanged and 0 otherwise, as indicated by the gray shading in Fig.~\ref{fig3}a; only the remaining part $\Delta\phi_\text{sy}^\text{int}$ is shown there.

\textit{Numerical convergence.} --- Within the no-jump evolution generated by $H^\mathrm{eff}$, the excitation number is conserved, so that the one- and two-excitation sectors are closed and are treated exactly, the latter in the complete $N(2N-1)$-dimensional basis. The only numerical parameter is therefore the tolerance of the time integration. All time evolutions are obtained with an adaptive eighth-order Runge--Kutta integrator (DOP853) at relative and absolute tolerances $\mathrm{rtol}=10^{-12}$ and $\mathrm{atol}=10^{-14}$. Tightening these to $\mathrm{rtol}=10^{-13}$ and $\mathrm{atol}=10^{-15}$ changes the survival probabilities, transfer probabilities and state overlaps by less than $2\times10^{-9}$, and the accumulated phases, including $\Delta\phi$, by less than $2\times10^{-8}$\,rad, for all parameter sets shown, including the edges of the parameter ranges of Figs.~\ref{fig2}c and~\ref{fig3} of the main text. The phase bound holds except in the immediate vicinity of the switching points of  $\Delta\phi^\mathrm{ex}$, where the reference amplitude vanishes. Repeating each propagation with $\Gamma\equiv0$, for which the evolution is unitary, yields a norm drift of at most $2\times10^{-9}$ over the full protocol.

\end{document}